\documentclass[%
reprint,
amsmath,
amssymb,
aps,
prb,
floatfix,
showkeys
]{revtex4-2}

\usepackage{graphicx} 
\usepackage{dcolumn} 
\usepackage{bm} 
\usepackage{xcolor}
\usepackage[colorlinks=true, allcolors=gray]{hyperref}
\usepackage{float}
\usepackage{verbatim}
\usepackage{amssymb} 

\makeatletter
\renewcommand{\p@subsection}{}
\renewcommand{\p@subsubsection}{}
\makeatother

\newcommand*{\cond}[0]{\vert}

\newcommand*{\fref}[1]{Fig.~\ref{#1}}
\newcommand*{\tref}[1]{Table~\ref{#1}}
\newcommand*{\eref}[1]{Eq.~\eqref{#1}}
\newcommand*{\sref}[1]{Section~\ref{#1}}
\newcommand*{\aref}[1]{Appendix~\ref{#1}}

\setcitestyle{authoryear,round}
\begin{document}

\title{
Uncertainty Quantification and Propagation in Atomistic Machine Learning}

\author{Jin Dai}
\author{Santosh Adhikari}
\author{Mingjian Wen}
\email{mjwen@uh.edu}
\affiliation{Department of Chemical and Biomolecular Engineering, University of Houston, Houston, TX, 77204, USA}

\date{\today}

\begin{abstract}

Machine learning (ML) offers promising new approaches to tackle complex problems and has been increasingly adopted in chemical and materials sciences.
Broadly speaking, ML models employ generic mathematical functions and attempt to learn essential physics and chemistry from a large amount of data.
Consequently, because of the limited physical or chemical principles in the functional form, the reliability of the predictions is oftentimes not guaranteed, particularly for data far out of distribution.
It is critical to quantify the uncertainty in model predictions and understand how the uncertainty propagates to downstream chemical and materials applications.
Herein, we review existing uncertainty quantification (UQ) and uncertainty propagation (UP) methods for atomistic ML under a united framework of probabilistic modeling. 
We first categorize the UQ methods, with the aim to elucidate the similarities and differences between them. 
We also discuss performance metrics to evaluate the accuracy, precision, calibration, and efficiency of the UQ methods and techniques for model recalibration.
With these metrics, we survey existing benchmark studies of the UQ methods using molecular and materials datasets. 
Furthermore, we discuss UP methods to propagate the uncertainty obtained from ML models in widely used materials and chemical simulation techniques, such as molecular dynamics and microkinetic modeling.
We also provide remarks on the challenges and future opportunities of UQ and UP in atomistic ML.

\end{abstract}


\maketitle

\section{Introduction}

Since the breakthrough in image recognition using deep neural networks (NNs) back in 2012 \citep{krizhevsky2012imagenent}, machine learning (ML) approaches have been increasingly leveraged to study complex chemical and materials systems.
They have achieved remarkable successes, from designing catalysts \citep{zahrt2019prediction, back2019toward}, to discovering functional materials \citep{axelrod2022learning, rao2022machine} and studying protein folding \citep{jumper2021highly, baek2021accurate}, to name a few.
The ML approaches applied in these tasks take advantage of a wide range of techniques, but they share a common core idea: modeling molecules, materials, and chemical reactions at the atomic scale and looking for patterns and trends in atomic data, which we refer to as \emph{atomistic machine learning}.

One of the most impactful developments in atomistic ML for chemical and materials science is creating interatomic potentials (i.e., force fields) to model the interactions between atoms.
This goes from early endeavors that model individual molecular/material systems (e.g., the feed-forward NN potential \citep{behler2007generalized, behler2021four} and Gaussian approximation potential (GAP) \citep{bartok2010gaussian, deringer2021gaussian}) to more recent efforts to build universal potentials for the entire periodic table (e.g., M3GNet \citep{chen2022universal}, CHGNet \citep{deng2023chgnet} and MACE-MP  
\citep{batatia2024foundation}).
While interatomic potentials remain an active focus of atomistic ML, 
the field has moved beyond and expanded to encompass the prediction of arbitrarily complicated molecular, materials, and reaction properties \citep{ceriotti2022beyond, redik2022extending}. 
These include molecular dipole moments \citep{unke2019physnet, gastegger2021machine}, bond strength \citep{john2020prediction, wen2021bondnet}, high-rank material tensors \citep{pakornchote2023straintensornet, wen2024equivariant}, neutron, x-ray, and vibrational spectroscopies \citep{chen2021machine, schienbein2023spectroscopy}, and reaction rates and yields \citep{heid2022machine, wen2023chemical}, among others.

Despite these successes, the reliability of atomistic ML models remains a significant concern \citep{peterson2017addressing, tavazza2021uncertainty, heid2023characterizing}. 
The complex and high-dimensional nature of chemical and materials systems, coupled with the limited availability of high-quality training data, can lead to models that are prone to overfitting, generalization errors, and poor transferability \citep{abdar2021review, wen2022improving, gawlikowski2023survey, psaros2023uncertainty}.
To address these challenges, it is imperative to quantify the uncertainty associated with model predictions, providing a measure of confidence in the results and helping to identify areas where the model may be unreliable.
The uncertainties can be broadly categorized into two types: aleatoric uncertainty and epistemic uncertainty \citep{hullermeier2021aleatoric, abdar2021review}.
Aleatoric uncertainty, also known as data uncertainty, arises from the inherent and irreducible noise in the data used for model development.
In atomistic ML, aleatoric uncertainty can come from, for example, the inexact exchange-correlation functional of the density function theory (DFT) employed to generate the training data \citep{wellendorff2012density, ruiz2005exchange, lejaeghere2016reproducibility, henkel2021uncertainty}.
Epistemic uncertainty, also referred to as model uncertainty, accounts for limitations of our knowledge or assumptions about a model. 
Factors such as model architecture, model parameters, and hyperparameter selection can all contribute to epistemic uncertainty.

Incorporating uncertainty quantification (UQ) into the ML model development process is a significant step forward; 
transmitting the uncertainty from the atomistic ML models to downstream tasks, known as uncertainty propagation (UP), is equally crucial \citep{wang2015combustion, honarmandi2019uncertainty, honarmandi2020uncertainty, abdi2024propagating}.
Atomistic ML models are typically trained to predict fundamental physical and chemical properties, such as the forces on atoms and chemical reaction rates. 
These properties are then used as inputs to other modeling techniques (analytical or numerical), such as molecular dynamics (MD) and microkinetic simulations, to obtain a final quantity of interest (QoI). 
The accurate and efficient propagation of uncertainty through the model chain to the QoI is essential to make informed decisions and assess the reliability of the final predictions.
Therefore, UQ and UP should be considered together when developing and applying atomistic ML models for chemical and materials applications.

Uncertainty analysis has been a crucial topic in chemical engineering and materials science. 
Classical methods such as polynomial chaos \citep{wiener1938homogeneous} have long been applied to study the uncertainty associated with the key parameters and outcomes in these domains \citep{phenix1998incorporation, reagan2005quantifying, villegas2012application}.
While uncertainty analysis traditionally serves to acknowledge model imperfections and has been primarily used for post-hoc analysis, with the emergence of ML approaches, the landscape has become increasingly rich and diverse. 
For example, uncertainty has become routinely used in active learning to select new atomic structures to enrich the dataset and subsequently retrain ML models. 
This has been demonstrated in both structure--property models \citep{gubaev2018machine, tian2021efficient, liu2022experimental} and MLIPs \citep{schwalbe2021differentiable, van2023hyperactive, zaverkin2024uncertainty}, among others.
Such proactive usage of uncertainty can greatly enhance the robustness of ML models and, additionally, it is data efficient.

This abundance of choices in ML-based uncertainty analysis, however, has created a significant challenge in decision-making for practitioners.
Several critical questions arise:
What are the fundamental similarities and differences between these UQ methods?
What constitutes good UQ methods, and how do their strengths and weaknesses compare in the context of atomistic ML?
Can the uncertainty obtained from an ML estimator be propagated to downstream chemical and materials applications, and if so, how?
Without clear answers to these questions, navigating the field of UQ and UP in atomistic ML can be a daunting task. 
It often leads to confusion, such as misinterpretation of the meaning and implications of the uncertainty, and difficulty in selecting an appropriate method for a given task. 

In this work, we provide a comprehensive review of selected, representative UQ and UP methods for atomistic ML, aiming to answer the above questions.
We assume the readers have a basic understanding of ML, but no prior knowledge of UQ and UP is required. 
We anticipate that this review will equip readers with a certain degree of certainty in navigating the space of uncertainty.
The paper is structured as follows. 
First, we present a primer on probabilistic modeling in \sref{sec:primer}, setting the stage for the coming sections. 
Next, in \sref{sec:uq}, we review and categorize the selected UQ methods, leveraging the concepts introduced in \sref{sec:primer}. 
In \sref{sec:eval}, we discuss ways to evaluate the quality of the UQ methods from four different perspectives:  accuracy, precision, calibration, and efficiency.
Additionally, in \sref{sec:up}, we explore UP techniques in widely used chemical and materials simulation techniques, using MD and microkinetic modeling as examples. 
Finally, we conclude by summarizing the current challenges of UQ and UP for atomistic ML and outlining opportunities for future research.
A summary of the most commonly used symbols is listed in \tref{tab:notation}, and a list of the abbreviations is provided in \aref{sec:abbr}.

\begin{table}[tbh!]
    \caption{Notation. Overview of the most commonly used symbols.}
    \label{tab:notation}
    \begin{ruledtabular}
    \begin{tabular}{ccc}
         Symbol & Explanation & Alternative \\
    \colrule
         $\theta$ & Model parameters&  \\ 
         $x$ & Model input &  \\ 
         $\hat y$ & Model output &  $\hat y = f(x; \theta)$ \\ 
         $X$ & Input set & $X = \{x_i\}_{i=1}^{N} $  \\ 
         $Y$ & Output set & $Y = \{y_i\}_{i=1}^{N} $  \\ 
         $D$ & Dataset & $D = (X, Y) $  \\ 
         $p(\theta)$ & Prior &  \\ 
         $p(Y \cond X, \theta)$ & Likelihood &  $p(D\cond\theta)$ \\ 
         $p(Y \cond X)$ & Marginal likelihood &   $p(D)$ \\ 
         $p(\theta \cond X, Y)$ & Posterior & $p(\theta \cond D$)\\ 
         $\mathcal{N}(\mu, \sigma^2)$ & Gaussian distribution & \\ 
         $\delta$  &  Uncertainty \\
         $\mathbb{E}_\theta [\cdot]$  & Expectation w.r.t.\ $\theta$ \\
    \end{tabular}
    \end{ruledtabular}
\end{table}

\section{Primer on probabilistic modeling}
\label{sec:primer}

In many atomistic ML problems, the goal is to obtain a regression model, $y = f(x; \theta)$, which maps the input $x$ (e.g., a chemical reaction) to the corresponding output $y$ (e.g., the reaction rate), where $\theta$ denotes all model parameters and can be determined from an observed dataset $D = \{(x_i, y_i)\}_{i=1}^N = (X, Y)$ consisting of $N$ data points.
In addition, we are interested in quantifying the uncertainty in the predicted outputs $y$. 

This section provides a brief review of the basic concepts in probabilistic modeling, laying the foundation for the discussion of uncertainty in subsequent sections. 
We first introduce \emph{Bayesian inference} to obtain the predictive distribution which expresses the uncertainty about the prediction $y$ for each input $x$.
Next, we discuss the frequentist \emph{maximum likelihood estimation} of model parameters $\theta$ and its link to the widely-used least-squares minimization.
Finally, we introduce \emph{evidence approximation}, a framework integrating frequentist estimates into the Bayesian approach to find approximate solutions.
Readers well-versed in probability theory may choose to skip this section and refer back to it as needed while reading the later sections.

\subsection{Distribution functions}
\label{sec:dist:fn}

\begin{figure*}[bth!]
    \centering
    \includegraphics[width=2\columnwidth]{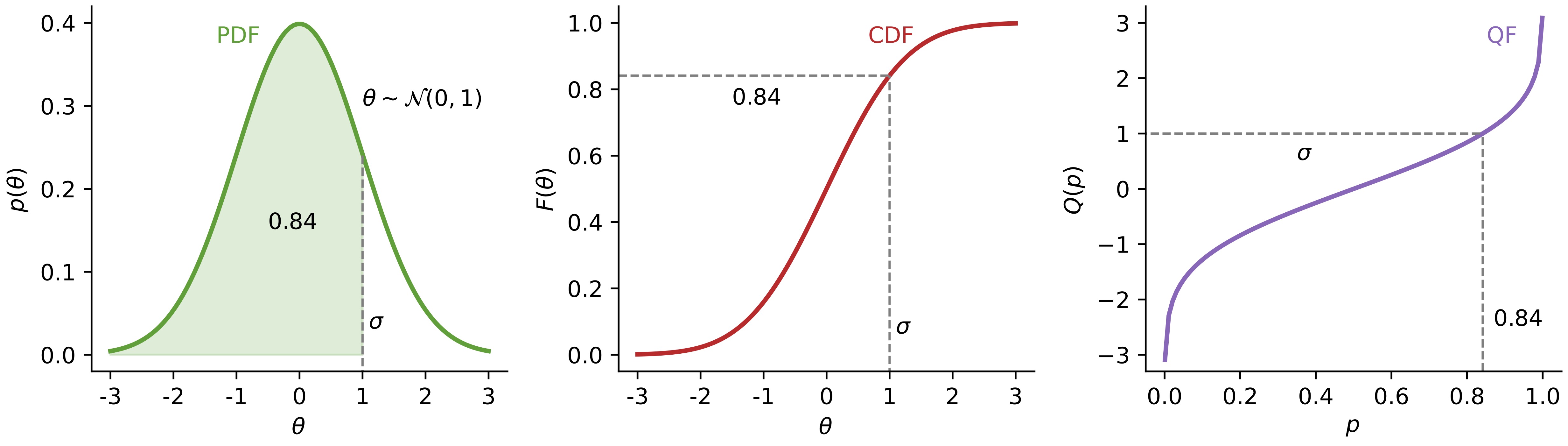}
    \caption{One-dimensional Gaussian distribution with a mean of $\mu=0$ and standard deviation $\sigma=1$.  
    The probability density function (PDF), cumulative probability function (CDF), and quantile function (QF). 
    The QF is the inverse of the CDF, as can be seen by switching the horizontal and vertical axes.}
    \label{fig:gaussian:dist}
\end{figure*}

A \emph{probability density function} (PDF), denoted as $p(\theta)$, is used to describe the probability distribution of a continuous random variable $\Theta$. 
The PDF represents the relative likelihood of the random variable taking on a specific value.
A \emph{cumulative distribution function} (CDF), denoted as $F(\theta)$, describes the probability that a random variable $\Theta$ takes a value less than or equal to $\theta$. 
The CDF can be obtained from the PDF via:
$F(\theta) = P(\Theta \leq \theta) = \int_{-\infty}^{\theta} p(t) dt$.  
A \emph{quantile function} (QF) maps probabilities to values of the random variable. 
It is defined as the inverse of the CDF, $Q(p) = F^{-1}(p)$, that is, for a given probability $p$, the quantile function returns the value $\theta$ 
such that its probability is less than or equal to an input probability value, i.e., $P(\Theta \leq \theta) = p$. 
As a concrete example, \fref{fig:gaussian:dist} presents the PDF, CDF, and QF of a one-dimensional standard Gaussian distribution $\mathcal{N}(0, 1)$ with a mean of 0 and standard deviation of 1. 

\subsection{Bayesian inference}
\label{sec:bayesian}

In the Bayesian view, probability provides a quantification of uncertainty \citep{gelman2013bayesian}.
Given an observed dataset $D$, we are interested in obtaining the conditional probability $p(y^* \cond x^*, D)$ of the output $y^*$ for a new input $x^*$.
From it, a statistical measure of the uncertainty in $y^*$ such as the variance can then be obtained.  
Given a parametric model, $y = f(x;\theta)$, this can be achieved in two steps.

First, obtaining the \emph{posterior distribution} over model parameters $\theta$ using Bayes' theorem \citep{gelman2013bayesian}:
\begin{equation} 
    \label{eq:bayes:theorem}
    p(\theta \cond D) = \frac{p(D\cond\theta) p(\theta)}{p(D)}.
\end{equation}
The \emph{prior distribution} $p(\theta)$ represents our prior information as to which parameters $\theta$ are likely to have generated the outputs before observing any data, based on previous knowledge, experience, or physical limitations.
The effects of the observed data $D$ come from the \emph{likelihood function}, $p(D\cond\theta)$, which quantifies the plausibility of $D$ for different realizations of $\theta$.
The denominator,   
\begin{equation} 
    \label{eq:marginal:likelihood}
    p(D) = \int p(D\cond\theta) p (\theta) \,\text{d}\theta,
\end{equation}
is called the \emph{marginal likelihood}, also known as the \emph{evidence} in the context of Bayesian statistics, which ensures that the posterior is a proper probability and thus integrates into one. 
It represents the likelihood of the observed data $D$, considering all possible values of the parameter $\theta$ weighted by their prior distribution.
Marginalization means evaluating this equation to obtain the marginal likelihood.
Bayes' theorem converts the prior probability over model parameters into the posterior probability by incorporating the evidence provided by the observed data.

Second, with the posterior over $\theta$, we can obtain the \emph{predictive distribution} for a new data point $(x^*, y^*)$ as 
\begin{equation} 
    \label{eq:predictive:dist}
    p(y^* \cond x^*, D) 
    = \int  p(y^* \cond x^*, \theta) p(\theta \cond D)  \,\text{d}\theta,
\end{equation}
where $p(y^* \cond x^*, \theta) $ is the likelihood given the new data point. 
It is related to the likelihood function for a dataset $p(D\cond\theta)$ in \eref{eq:bayes:theorem}, and further discussion on this will be provided in \sref{sec:mle}. 
From the predictive distribution, we can readily obtain, e.g., the mean as the final prediction and the variance as a point estimate of the uncertainty.

Although theoretically sound, a major practical limiting factor of the full Bayesian approach lies in the complexity of evaluating the posterior.
In particular, the marginal likelihood in \eref{eq:marginal:likelihood} can only be analytically evaluated for simple models like linear regression. 
Numerical techniques such as sampling methods and variational inference have to be undertaken to evaluate the predictive distribution 
for more complicated models \citep{bishop2006pattern}.    

One can sample the posterior distribution using Monte Carlo (MC) methods, such as Markov chain Monte Carlo (MCMC), and then obtain the predictive mean and uncertainty \citep{neal1993probabilistic, neal2003slice}.
Sampling methods are accurate, flexible, and can be applied to a wide range of models. 
However, they are still computationally intensive because a large number of samples might be needed for convergence; thus, they are mainly used for small-scale problems \citep{bishop2006pattern}.

Alternatively, the variational inference approach tackles the challenge by employing another distribution $q(\theta)$ to approximate the true posterior $p(\theta\cond D)$, and then using $q(\theta)$ to evaluate the predictive distribution in \eref{eq:predictive:dist}.
The approximate distribution $q(\theta)$ is typically much simpler, and it is optimized to resemble the true posterior, e.g., by minimizing the Kullback--Leibler divergence \citep{kullback1951information,mackay1992practical} between $q(\theta)$ and $p(\theta\cond D)$.
Although not exact, variational inference offers a computationally efficient approach to evaluate the predictive distribution.
The MC dropout method to obtain uncertainty in NN models proposed by \cite{gal2016uncertainty} adopts this approach, and it will be further discussed in \sref{sec:dropout}.

\subsection{Maximum likelihood}   
\label{sec:mle}

Maximum likelihood estimation (MLE) is a frequentist approach to estimate the optimal parameters $\theta$ of a model $y = f(x; \theta)$. 
It is equivalent to the least-squares parameter optimization technique widely used in science and engineering.
MLE provides a point estimate of the parameters and thus predictive uncertainty cannot be directly quantified from a model trained using MLE alone. 
However, MLE serves as a fundamental concept in parameter estimation and forms the basis of many other UQ methods.

In MLE, we focus on the likelihood function $p(D\cond\theta)$, and do not care about the prior and posterior (see \eref{eq:bayes:theorem}).
We assume that the observed output $y$ is given by the model prediction $f(x; \theta)$ with an additive error $\epsilon$, i.e., $y = f(x; \theta) + \epsilon$.
A Gaussian distribution with zero mean is a reasonable choice for the error, $p(\epsilon)  = \mathcal{N} (\epsilon\vert 0,\sigma^2) $, where $\sigma^2$ is the variance.
This is equivalent to the Gaussian distribution in which $y$ is regarded as the random variable with the model prediction $\hat y = f(x; \theta)$ as its mean and $\sigma^2$ as the variance:
\begin{equation}  
    \label{eq:likelihood:single}
    p(y|x, \theta)
    = \mathcal{N}(y \cond \hat y, \sigma^2)
    = \frac{1}{\sqrt{2\pi \sigma^2}} \exp\left(-\frac{(y - \hat y)^2}{2 \sigma^2}\right).
\end{equation}
In other words, this is the likelihood of $\theta$ for a single data point $(x, y)$.
Now consider the observed dataset $D$ where each data point is drawn independently from the distribution in \eref{eq:likelihood:single}. 
We can obtain the likelihood function for the dataset as the product of the likelihood for each data point:
\begin{equation} 
    \label{eq:likelihood:dataset}
    p(D \cond \theta) 
    = p(Y \cond X,\theta) 
    = \prod_{i=1}^{N} p(y_i \cond x_i, \theta).
\end{equation}
The optimal model parameters can thus be obtained by maximizing \eref{eq:likelihood:dataset} with respect to (w.r.t.) $\theta$, 
which is equivalent to minimizing the negative log-likelihood (NLL): 
\begin{equation} 
    \label{eq:NLL}
    \text{NLL} = -\log p(D\cond\theta)=\frac{1}{2} \sum_{i=1}^N \left(\log(2\pi \sigma^2) + \frac{(y_i - \hat y_i)^2}{\sigma^2} \right),
\end{equation}
because the logarithm function is monotonically increasing.
This transformation converts a product of probabilities into a sum of log probabilities, which is often more convenient to optimize.

We note that in MLE, the variance $\sigma^2$ is modeled as a single constant, albeit unknown. 
Consequently, minimizing the NLL is equivalent to the familiar least-squares minimization w.r.t.\ $\theta$ using the loss: 
\begin{equation} 
    \label{eq:least:squares}
    L(\theta) = \frac{1}{2} \sum_{i=1}^{N}{(y_i - \hat y_i)^2}.
\end{equation}
Recall that the model parameters $\theta$ are implicitly indicated in the model prediction $\hat y = f(x; \theta)$.
With the optimal point estimate of the parameters, $\theta_\text{opt}$, we can then get model prediction as $y^* = f(x^*; \theta_\text{opt})$ for any new input $x^*$.
Again, we note that no information on uncertainty can be directly obtained from this approach only.

\subsection{Evidence approximation} 
\label{sec:evidence:approx}
  
While the full Bayesian approach can produce predictive uncertainty, it can be computationally demanding to obtain.
On the other hand, it is straightforward to get a point estimate of the optimal model parameters using MLE, but the uncertainty cannot be quantified. 
The \emph{evidence approximation} approach \citep{gull1989developments, mackay1992bayesian}, also known as the \emph{empirical Bayes} \citep{bernardo2009bayesian}, \emph{generalized maximum likelihood} \citep{wahba1985comparison}, or \emph{type 2 maximum likelihood} \citep{berger1985statistical}, lies between the two extremes.

Evidence approximation is a method that looks at how well a model fits the data overall. 
It focuses on the marginal likelihood in \eref{eq:marginal:likelihood}, which provides the \emph{model evidence} of observing the data marginalized over the parameters, meaning that it calculates the probability of seeing the observed data under all possible parameter values of the model.
In evidence approximation, the prior $p(\theta)$ in \eref{eq:marginal:likelihood} is further parameterized using a set of hyperparameters $\xi$, becoming $p(\theta\cond\xi)$. 
Consequently, the marginal likelihood is \citep{bishop2006pattern}:
\begin{equation} 
    \label{eq:model:evidence}
    p(D\cond\xi) = \int p(D\cond\theta,\xi) p (\theta\cond\xi) \,\text{d}\theta.
\end{equation}
The introduction of the hyperparameters $\xi$ increases the model's capacity, meaning that the model has increased flexibility to capture the underlying structure present in the data.
In practice, the prior distribution $p(\theta\cond\xi)$ is often chosen to be conjugate to the likelihood $p(D\cond\theta,\xi)$. 
In other words, $p(\theta\cond\xi)$ is specifically selected to match the form of the likelihood $p(D\cond\theta,\xi)$ such that the integration in \eref{eq:model:evidence} has a closed-form solution, thereby simplifying the process to obtain $p(D\cond\xi)$.
For example, if the likelihood is the Binomial distribution, a conjugate prior for it is the Beta distribution, and then \eref{eq:model:evidence} can be analytically evaluated to obtain the model evidence, which is a Beta distribution as well \citep{degroot2012probability}.

In a full Bayesian setting, after obtaining $p(D\cond\xi)$, one would also marginalize over the hyperparameters $\xi$ to obtain $p(D)$ and then perform Bayesian inference.
However, this marginalization typically does not have an analytical solution.
Instead, in evidence approximation, we get a point estimate of $\xi$ by maximizing the model evidence $p(D\cond\xi)$ w.r.t.\ $\xi$.
Then, the model evidence, prior, and likelihood are all evaluated using the optimal hyperparameters $\xi_\text{opt}$ to perform Bayesian inference using Eqs.~\eqref{eq:bayes:theorem} and \eqref{eq:predictive:dist}.

Alternatively, one can obtain the predictive uncertainty directly from $p(\theta\cond\xi_\text{opt})$, provided that the prior is chosen to be of a specific form, e.g., as a high-order distribution on top of the likelihood. 
This is the \emph{evidential regression} approach recently proposed by \cite{amini2020deep}.
In \sref{sec:deep:evidential}, we will further discuss this approach and explain how to get the uncertainty.

\section{Uncertainty Quantification}
\label{sec:uq}

\begin{figure*}[bt!]    
    \centering     
    \includegraphics[width=2\columnwidth]{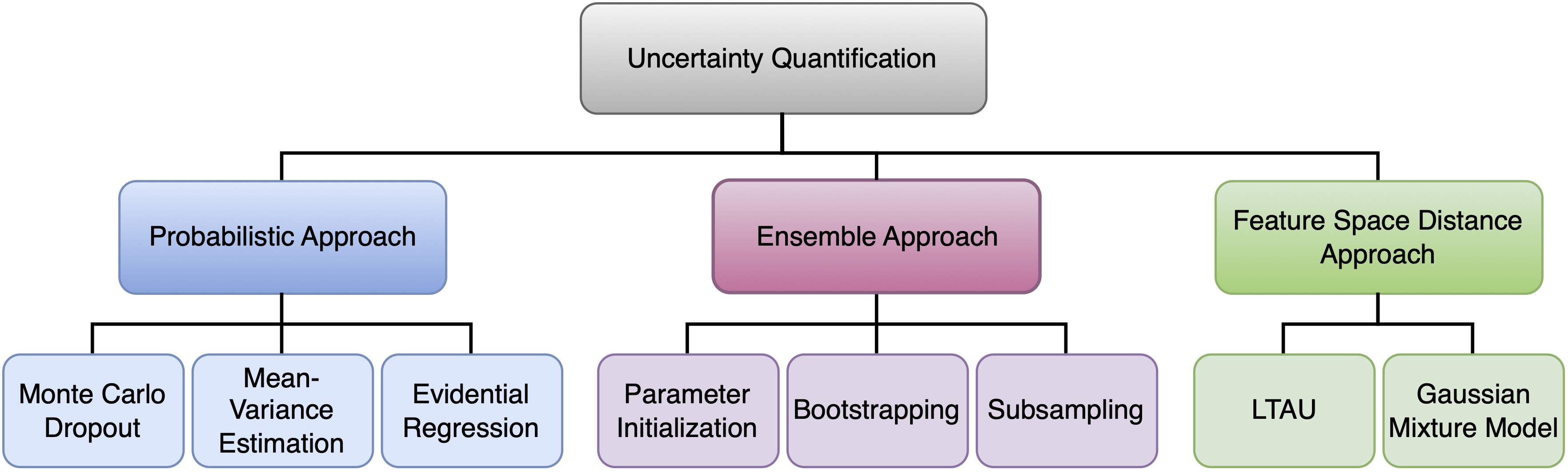}
    \caption{Categorization of uncertainty quantification methods in atomistic machine learning. LTAU: loss trajectory analysis for uncertainty.}    
\label{fig:uq:categorization}
\end{figure*}

A large number of UQ methods have been developed for ML models.  
Here, we discuss several selected, representative ones for atomistic ML for chemical and materials applications; 
in particular, we focus on UQ methods for NN models.
We classify them into three categories (\fref{fig:uq:categorization}) mainly based on the model construction strategy: 
probabilistic approach, ensemble approach, and feature space distance approach.
A probabilistic approach models some distribution discussed in \sref{sec:primer} and derives uncertainty from it.
An ensemble approach builds multiple models and obtains the variance in the predictions as the uncertainty.
A feature space distance approach measures some ``distance'' of a data point to the model training data and regards the distance as the uncertainty.
Although each method is placed under a single category in \fref{fig:uq:categorization}, some can belong to different categories.
For example, MC dropout can also be regarded as an ensemble approach.
We note that there are other ways to categorize the UQ methods, such as the one based on model utilization strategy \citep{gawlikowski2023survey}.

In this section, we discuss the UQ methods, examining how a model is trained and how uncertainty is obtained.
In addition, we provide example applications for chemical and materials' problems.
A summary of the UQ methods is provided in \tref{tab:summary}.

\begin{table*}[tbh!]
\caption{Summary of the UQ methods.
Category denotes the class to which a method belongs.  
For the probabilistic methods, we provide the specific type of probabilistic approach the method belongs to.
UQ measure denotes the quantity being used as the uncertainty.
Efficiency means the number of models that need to be optimized in the training stage and the number of model evaluations that need to be performed to obtain the uncertainty in the inference stage.
A method with a check mark indicates that it can be used to conduct sampling-based UP.
LTAU: loss trajectory analysis for uncertainty; NLL: negative log-likelihood.}
\label{tab:summary}
\begin{ruledtabular}
\begin{tabular}{ccccc}
                               & Category            & UQ measure & 
                               Efficiency   & Sampling-based \\
                               &                     &              &(training/inference)  
                               & UP \\
 \colrule                               
Monte Carlo dropout            & Bayesian            & Variance     & One/Multiple         & \checkmark  \\
Mean-variance estimation       & Maximum likelihood                 & Variance     & One/One                         \\
Evidential regression          & Evidence approximation  & Variance     & One/One                         \\
Parameter initialization       & Ensemble            & Variance     & Multiple/Multiple    & \checkmark  \\
Bootstrapping                  & Ensemble            & Variance     & Multiple/Multiple    & \checkmark  \\
Subsampling                    & Ensemble            & Variance     & Multiple/Multiple    & \checkmark  \\
LTAU                           & Feature space distance   & Error ratio              & One/One                         \\
Gaussian mixture model         & Feature space distance   & NLL          & Two/One                          
\end{tabular}
\end{ruledtabular}
\end{table*}

\subsection{Probabilistic approach}

\subsubsection{Monte Carlo dropout} 
\label{sec:dropout}

The dropout technique was originally proposed by \cite{srivastava2014dropout} as a regularization technique to alleviate overfitting in deep NN models. 
It was adapted by \cite{gal2016dropout} to approximate the Bayesian approach for UQ mentioned in \sref{sec:bayesian}.
Their method, known as MC dropout, can be theoretically viewed as sampling from a Bernoulli prior distribution over the weights of the NN, and then taking advantage of the variational inference technique to approximate the posterior distribution.
Practically, dropout is used at both training and inference time, allowing the model to estimate uncertainty by considering multiple predictions from different subsets of the network. 

The model can be trained by minimizing a loss between its predictions and the corresponding reference values to obtain the optimal NN parameters.
At each training step, dropout randomly sets the outputs of a fraction of the nodes in an NN to zero (e.g., the dashed nodes in \fref{fig:prob:uq}a), effectively creating an ensemble of thinned sub-networks. 
Once trained, we use the model in a similar way.
Multiple forward passes are performed through the NN, each with different nodes being dropped, to obtain multiple predictions.
The predictions are then averaged to obtain the final prediction and their variance is computed as the predictive uncertainty.
In this sense, MC dropout can also be thought of as a frequentist ensemble approach to be discussed in \sref{sec:ensemble}.

It is important to note that MC dropout is an approximation and may not capture the full posterior distribution over the NN's weights. 
Nevertheless, it has gained popularity as a practical technique in atomistic ML.
For example, \cite{wen2020uncertainty} have developed a dropout NN model for carbon allotropes and shown that the obtained uncertainty can reliably distinguish diamonds from graphene and graphite.

\subsubsection{Mean-variance estimation}

\begin{figure*}[tbh!]
    \centering    
    \includegraphics[width=2\columnwidth]{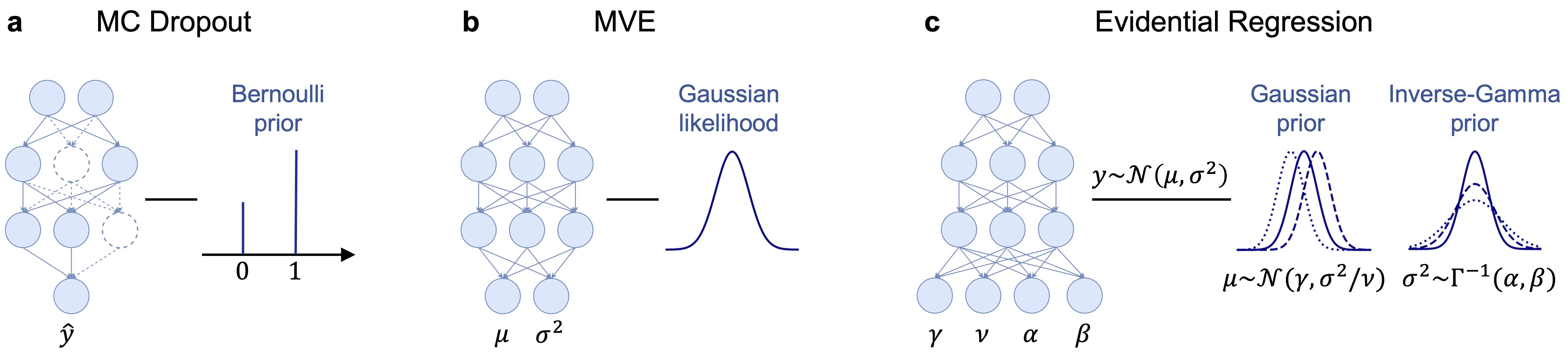}
    \caption{Schematic illustration of the probabilistic UQ approaches: 
    (a) MC dropout; (b) MVE; and (c) Evidential regression.}   
    \label{fig:prob:uq}
\end{figure*}

The mean-variance estimation (MVE) method, first introduced by \cite{nix1994estimating}, enables the use of a single deterministic NN to obtain the predictive uncertainty.
This method largely follows the MLE framework (\sref{sec:mle}) but with slight adjustments. 
In MLE, the observed data are assumed to be independent and identically distributed (i.i.d.) samples from a Gaussian distribution, where the mean $\mu$ is given by a parameterized model $ \hat y = f(x; \theta)$ of the input $x$, while the same constant variance $\sigma^2$ is assumed for all observed data (see \eref{eq:likelihood:single}).
In contrast, MVE uses different variances for different data points to model the uncertainty. 
In other words, the observed data are assumed to be drawn from the Gaussian: $\mathcal{N}(\mu (x), \sigma^2(x))$,
in which both the mean $\mu$ and the variance $\sigma^2$ are parameterized models of the input $x$.
In practice, an NN with two output nodes can be employed as the parameterized model, one node predicting the mean $\mu$ and the other for the variance $\sigma^2$ (\fref{fig:prob:uq}b).

The training process involves using MLE to optimize the NN's parameters. 
In this case, since the variance is not a constant, MLE is not equivalent to least-squares minimization with the loss in \eref{eq:least:squares} anymore.
Instead, we will need to directly minimize the NLL in \eref{eq:NLL}.
Once trained, the predicted mean $\mu$ by the NN gives the final prediction, and the predicted variance $\sigma^2$ serves as the uncertainty.

MVE has been adopted by \cite{tan2023single} to predict the energies of small molecules, among others. 
They found that it has the highest average test error when compared to other methods and suggested that this might be attributed to the harder-to-optimize NLL loss function, which has been reported in \cite{seitzer2022on}.

\subsubsection{Evidential regression}
\label{sec:deep:evidential}

The deep evidential regression method \citep{amini2020deep} adopts the evidence approximation approach discussed in \sref{sec:evidence:approx} to provide uncertainty for NN models. 
\eref{eq:model:evidence} gives the model evidence for an entire dataset.
To simplify the discussion, here we focus on the model evidence for a single observation $(x, y)$,
\begin{equation} 
    \label{eq:evidence:data:point}
    p(y\cond\xi) = \int p(y\cond\theta,\xi) p (\theta\cond\xi) \,\text{d}\theta,
\end{equation}
where we omit the conditional dependence on $x$ for simplicity.
Similar to MLE, the data is assumed to be sampled from a Gaussian likelihood:
\begin{equation} 
    \label{eq:evidence:likelihood}
    p(y\cond\theta,\xi) = \mathcal{N}(\mu, \sigma^2), 
\end{equation}
where $\theta = (\mu, \sigma^2)$, denoting the mean and variance of the Gaussian. 
However, different from MLE where the mean and variance are fixed constants, here, the mean and variance are parameterized over $\xi$.
\cite{amini2020deep} proposed to parameterize $\mu$ and $\sigma^2$ using the Gaussian and Inverse-Gamma distributions, respectively:
\begin{equation}
   \mu \sim \mathcal{N}(\gamma, \sigma^2\nu^{-1}) 
   \quad\quad 
   \sigma^2 \sim  \Gamma^{-1}(\alpha, \beta),
\end{equation}
where $\xi = (\gamma, \nu, \alpha, \beta)$ denotes the four hyperparameters.
Further, it is assumed that the prior $p(\theta\cond\xi)$ can be factorized into the product of the distributions of $\mu$ and $\sigma^2$, then it can be written as, 
\begin{equation} 
    \label{eq:nig}
    p(\theta\cond\xi) = 
   \mathcal{N}(\gamma, \sigma^2\nu^{-1})  
   \cdot
   \Gamma^{-1}(\alpha, \beta),
\end{equation}
which is called the \emph{Normal Inverse-Gamma} (NIG) distribution, a high-order evidential distribution.  
Sampling from NIG yields instances of lower-order likelihood functions from which the data is drawn.

The NIG prior in \eref{eq:nig} is the conjugate distribution to the Gaussian likelihood in \eref{eq:evidence:likelihood};
therefore, the model evidence in \eref{eq:evidence:data:point} can be evaluated analytically, resulting in the Student-t distribution:
\begin{equation} 
    \label{eq:student:t}
   p(y\cond\xi) = \text{St} \left(y; \gamma, \frac{\beta(1+\nu)}{\nu\alpha}, 2\alpha  \right).
\end{equation}

Recall from \sref{sec:evidence:approx} that the optimal hyperparameters $\xi = (\gamma, \nu, \alpha, \beta)$ are obtained by maximizing the model evidence, namely \eref{eq:student:t}.
In the deep evidential regression method by \cite{amini2020deep}, the hyperparameters are further parameterized by an NN and obtained as the output of the NN with four output nodes, one for each hyperparameter (\fref{fig:prob:uq}c).
So, instead of $\xi$, the parameters in the NN are optimized.   
In practice, we do not maximize \eref{eq:student:t} but minimize the equivalent NLL of \eref{eq:student:t} for numerical stability.
In addition, extra regularization terms can be added to remove misleading evidence.  
We refer to \cite{amini2020deep} for the technical details of model training. 

Once trained, the final prediction can be computed from the NIG as \citep{amini2020deep}
\begin{equation}
    \mathbb{E}[\mu] = \gamma,
\end{equation}
and the uncertainty as 
\begin{equation}
    \mathbb{E}[\sigma^2] + \text{Var}[\mu]
    = \frac{\beta}{\alpha- 1} + \frac{\beta}{\nu(\alpha - 1)}.
\end{equation}

Deep evidential regression has been used by \cite{soleimany2021evidential}
to predict molecular properties, and the obtained uncertainty has been successfully used to achieve sample-efficient training of property estimator and to guide the virtual screening for antibiotic discovery.
\cite{gruich2023clarifying} have also demonstrated its effectiveness in heterogeneous catalysis applications.

\subsection{Ensemble approach}
\label{sec:ensemble}

\begin{figure*}[tbh!]
    \centering    
    \includegraphics[width=2\columnwidth]{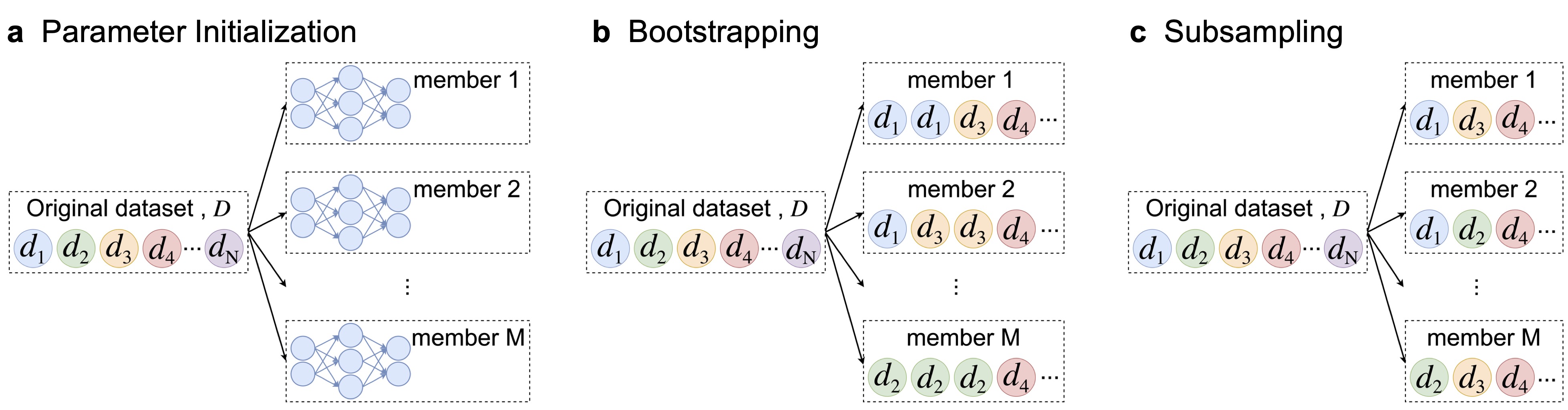}
    \caption{Schematic illustration of the ensemble UQ approaches: (a) parameter initialization, (b) bootstrapping, and (c) subsampling. }
    \label{fig:ensemble}    
\end{figure*}
The ensemble approach, characterized by its simplicity and diverse construction methods, combines multiple models to create a more robust predictive model, surpassing individual model limitations \citep{zhou2012ensemble}.
The frequentist ensemble approach is easy to implement and can be applied to a large number of regression algorithms. 
It might be computationally expensive when compared to other UQ methods, but they can be naively paralleled.
A couple of methods exist to construct an ensemble, such as parameter initialization, bootstrapping, and subsampling (\fref{fig:ensemble}).

The first type of method involves fitting models with different parameter initializations to the same dataset.
\cite{lakshminarayanan2017simple} proposed the use of ensembles for estimating the uncertainty of NNs. 
NNs of the same structure are created, but their parameters are initialized to be different. 
Each member of the ensemble is trained on the entire training set, meaning that all members have access to the same data. 
 
The second type of method focuses on fitting the same model to different datasets. 
Bootstrapping is a widely used method to generate multiple derived datasets from a given dataset.
Each subset, called a bootstrap sample, is created by randomly sampling the same number of data points from the original dataset with replacement \citep{hastie2009elements}.
This means that each data point has an equal probability of being selected, and the same data point can be included multiple times in the bootstrap sample.
This process is repeated $M$ times, resulting in $M$ bootstrap samples.
Then, $M$ models are trained separately, each using one of the bootstrap samples. 

Subsampling \citep{politis1994large, Politis_1999_Weak_convergence} is an alternative to bootstrapping to create multiple derived datasets from a given dataset.
It is similar to bootstrapping but with a key difference: subsampling is performed without replacement and thus each data point can only appear at most once in each subset.
As a result, each sample consists of fewer data points than the original dataset.

The final ensemble prediction and uncertainty are obtained by combining the outputs  $\hat y_1, \hat y_2, \dots, \hat y_M$ of all members.
The mean,
\begin{equation}
\bar y = \frac{1}{M} \sum_{i=1}^M{\hat y_i},
\end{equation}
gives the final prediction, and the variance,
\begin{equation} 
\sigma^2 = \frac{1}{M-1} \sum_{i=1}^M (\hat y_i - \bar y) ^2,
\end{equation}
gives the predictive uncertainty.

\subsection{Feature space distance approach} 

Another category of UQ approach is based on some distance measure in the model's feature/latent space.
It assumes that data points resembling each other are positioned closer to one another in the feature space.
Therefore, for a given test data point, if it is close to the training data in the feature space, the predictive uncertainty is low; otherwise, the predictive uncertainty is high.

In this approach, the uncertainty is typically obtained in a two-step process. 
First, given a primary model like an NN, a point estimate of the optimal model parameters is obtained using a training technique such as MLE. 
Second, construct another model to measure the distance between a test data point and the training data in the primary model's feature space. 
When dealing with NNs, the feature space can be chosen as the last but one layer or other internal layers that are considered suitable.

\begin{figure}[bth!]
    \centering
    \includegraphics[width=1\columnwidth]{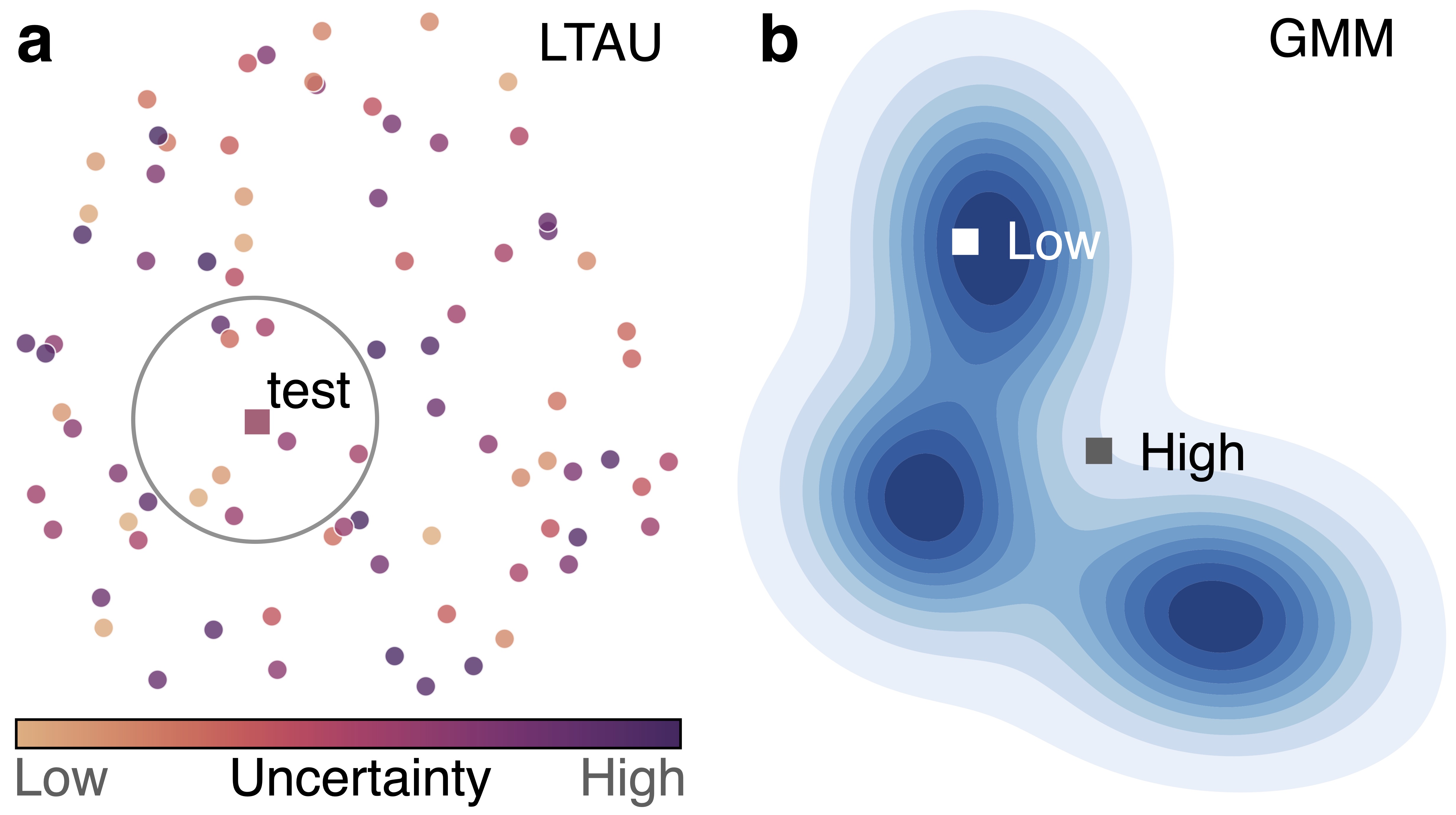}
    \caption{Schematic illustration of the UQ approaches based on feature space distance. 
    (a) LTAU assigns the average uncertainty of neighboring training data (circles) to a test point (square). 
    (b) GMM models the density of the training data in the feature space, and a test point has low uncertainty if it is located in a dense region.
    }
    \label{fig:feat:dist}
\end{figure}

\subsubsection{LTAU}

Loss trajectory analysis for uncertainty (LTAU) measures the distance in the Euclidean space.
It begins by training an NN model to predict an atomic property $y$, chosen to be the forces on atoms in \cite{vita2024ltau}.
During training, besides optimizing the model parameters, the error $\epsilon_i = \| y_i - \hat y_i \|^2$ between the model prediction $\hat y_i$ and its corresponding reference $y_i$ for each atom $i$ is recorded as 
\begin{equation} 
\label{eq:ltau:error:traj}
  T_i = \{\epsilon_i^1, \epsilon_i^2, ..., \epsilon_i^E \} ,
\end{equation}
where the super index denotes the training epoch at which the error is logged, and $E$ is the total number of epochs to train the model.

After training, we get a set of errors $T_i$ along the loss trajectory for each data point $i$ and then convert the errors to the model's confidence score for that data point via 
\begin{equation} 
\label{eq:ltau:confidence}
 p_i = P(\epsilon_i^e \in T_i \leq atol) , 
\end{equation}
which means the ratio of data points in $T_i$ whose value is smaller than or equal to a tolerance $atol$.
A reasonable choice for $atol$ would be the mean absolute error (MAE) between the predictions from the trained primary model and their references.
The value of $p_i$ is in the range of $[0, 1]$.
The uncertainty of each data point is calculated as 
\begin{equation}
   \delta_i  = 1 - p_i,
\end{equation}
which can be interpreted as the probability that the model's prediction on data point $i$ will have an error larger than the MAE.

The uncertainty for a new data point, $j$, is obtained by averaging the uncertainties of its nearest $K$ neighbors in the training data (\fref{fig:feat:dist}a):
\begin{equation}
 \delta_j = \frac{1}{K} \sum_{i \in N_j} \delta_i ,
\end{equation}
where $N_j$ denotes the set of neighbors.
The nearest neighbors can be determined by performing a similarity search based on Euclidean distance in the feature space.

LTAU has been successfully applied to tune the training–validation gap in NN potentials for carbon materials and predict the errors in relaxation trajectories of catalysts \citep{vita2024ltau}.

\subsubsection{Gaussian mixture model}

An alternative to Euclidean distance is measuring the density. 
If a test point is in the dense region where the training data are located in the feature space, then this test point has low uncertainty (\fref{fig:feat:dist}b).
The density can be estimated by a Gaussian mixture model (GMM).

After training the primary model such as an NN, we get a set of feature vectors $H = \{h_1, h_2, \dots, h_N \}$, each representing a training data point in the feature space.
The set of feature vectors is then used to train the second GMM model.
We aim to capture the underlying structure of the feature vectors using a GMM:
\begin{equation} \label{eq:gmm}
 p(h_i|w, \mu, \Sigma)=\sum_{m=1}^{M} w_m \mathcal{N}(h_i | \mu_m, \Sigma_m),   
\end{equation}
which is linear combination of $M$ Gaussian functions of respective mean $\mu_m$ and covariance $\Sigma_m$, with weights $w_m$. 
Each Gaussian function is a multidimensional distribution in the feature space, and thus $\mu_m$ is a vector and $\Sigma_m$ is a matrix.
We assume a dataset consists of i.i.d.\ samples from this GMM likelihood.
Then, the NLL for the dataset can be written as
\begin{equation}
\text{NLL}(H | w, \mu, \Sigma) = -\sum_{i=1}^{N} \log \left( \sum_{m=1}^{M} w_m \mathcal{N}(h_i | \mu_m, \Sigma_m) \right),
\end{equation}
which can be derived in the same way as from \eref{eq:likelihood:single} to \eref{eq:NLL} for MLE.
To train the GMM, we minimize the NLL w.r.t.\ $\mu$, $\Sigma$, and $w$, meaning that we adjust the location and shape of the Gaussian distributions, as well as the weight of each member such that the GMM model best describes the density of the training data in the features space.
The optimization can be performed via a gradient-based technique, or, more typically, using the expectation-maximization algorithm \citep{hastie2009elements}.

Once the GMM model is trained, the uncertainty for a new data point $x^*$ can be obtained in two steps.  
First, obtain its feature vector $h^*$ using the primary model. 
Second, compute its NLL, 
\begin{equation}
\text{NLL}(h^* | w, \mu, \Sigma) = -\log \left( \sum_{m=1}^{M} w_m \mathcal{N}(h^* | \mu_m, \Sigma_m) \right),
\end{equation}
and the NLL can be regarded as the predictive uncertainty \citep{zhu2023fast}.

The GMM method has been employed by \cite{zhu2023fast} to build NN interatomic potentials, leveraging the uncertainty estimates for active learning and efficient training data selection.

Overall, feature space distance approaches, such as LTAU and GMM, measure how far a data point is from the distribution of the training data, essentially indicating whether the point is out of distribution.
The predicted uncertainty, however, does not necessarily scale with the prediction error; 
therefore, recalibration of the uncertainty is typically needed if one intends to use the uncertainty as a proxy of the prediction error.
Demonstration of the quality of the approach is provided in \sref{sec:eg}, and further discussion of the recalibration is given in \sref{sec:recalibration}.

\section{Performance Evaluation}
\label{sec:eval}

\begin{figure}[h]
    \centering
    \includegraphics[width=.7\columnwidth]{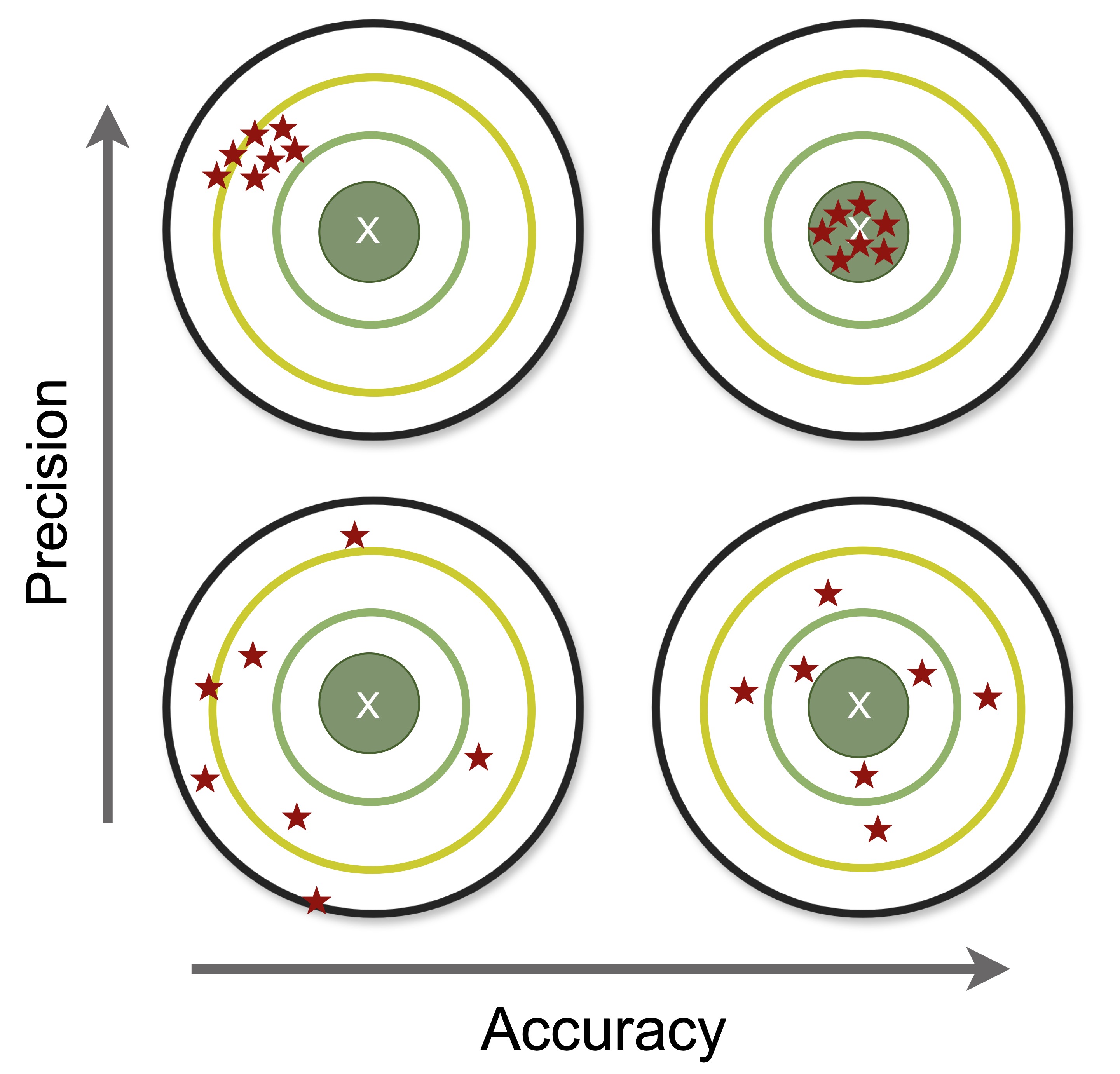}
    \caption{Uncertainty and prediction error.
    Loosely speaking, uncertainty gives the precision of the predictions, meaning how tightly the predictions are distributed against each other, 
    while prediction error measures the accuracy of the predictions, meaning the distance between the prediction and the true value.
    }
    \label{fig:acc:prec}
\end{figure}

What makes a UQ method effective?
Uncertainty is solely a property of model predictions, providing information about their \emph{precision}---the degree to which the predicted values are concentrated around each other (\fref{fig:acc:prec}). 
However, uncertainty does not directly measure the \emph{accuracy} of the predictions, i.e., how close they are to the true observations.
Despite this, a key application of uncertainty is to use it as an indicator of the likely accuracy of the predictions. 
Ideally, a prediction with high uncertainty should indicate a large error and thus be less reliable. 
The degree to which the uncertainty aligns with the accuracy is called \emph{calibration}.

An effective UQ method should be accurate, precise, and well-calibrated;
in addition, it should be computationally efficient for practical usage. 
These four aspects evaluate UQ methods from different perspectives.
In this section, we discuss performance evaluation for UQ methods, focusing on uncertainty calibration, a concept that, we believe, may be less familiar to researchers working on atomistic ML. 
We also examine scoring metrics for UQ evaluation, comment on the pros and cons of existing UQ methods, and provide concrete examples by drawing insights from existing benchmark studies. 
Furthermore, we introduce recalibration techniques to improve UQ performance.

\subsection{Calibration}

Calibration measures the statistical consistency between the predictions and observations, a property that depends on both the predictions and observations \citep{gneiting2007probabilistic}.
Uncertainty calibration has been extensively studied in the context of classification problems. 
Perfect calibration in this setting means that the confidence assigned to a class equals the probability of the prediction belonging to that class \citep{guo2017calibration,scalia2020evaluating}. For instance, if we have 10 predictions, each with a confidence of 0.8, we expect 8 of them to be correctly classified.

Uncertainty calibration for regression is less intuitive because the model predicts continuous values, rather than discrete labels as in classification. 
Nevertheless, following the groundbreaking work by \cite{gneiting2007probabilistic}, methods that extend the uncertainty calibration approach for classification have been proposed for regression problems \citep{kuleshov2018accurate, levi2022evaluating} and adopted in atomistic ML. 
Here, we discuss two such methods: \emph{interval based} and \emph{error based} regression uncertainty calibration.
As a technical note, we will use $\delta$ to denote uncertainty in general. 
However, for some models, uncertainty is represented by the variance $\sigma^2$ (see \tref{tab:summary}). 
Therefore, these two notations will be used interchangeably when appropriate.

\begin{figure}
    \centering
    \includegraphics[width=0.9\columnwidth]{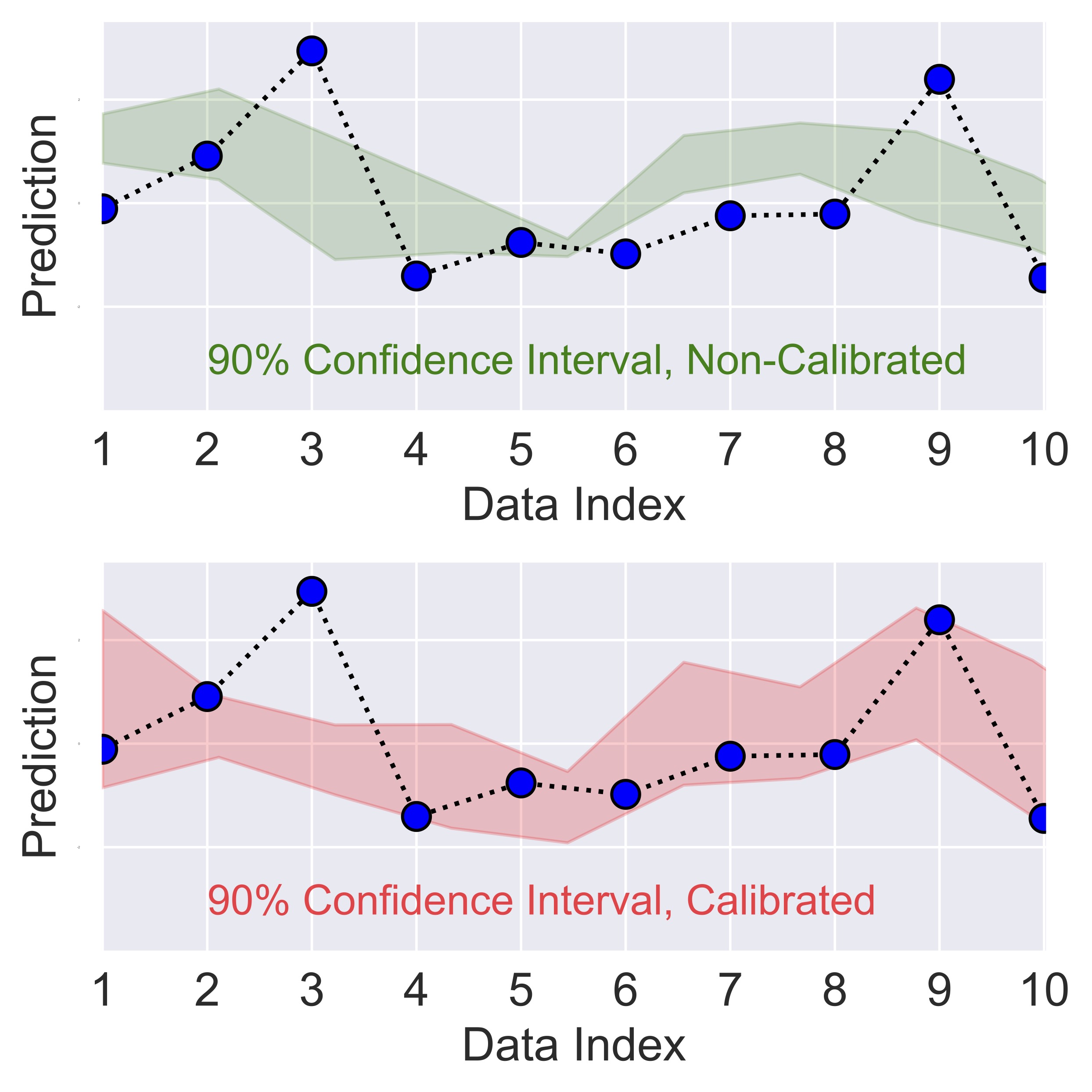}
    \caption{Illustration of a non-calibrated regression model (top) and a calibrated one (bottom). 
    The plot is inspired by \cite{kuleshov2018accurate} but generated using arbitrary data.}
    \label{fig:calib}
\end{figure}

\subsubsection{Interval based approach}
\label{sec:calib:interval}

Loosely speaking, in a regression setting, calibration means that a model prediction should fall in a given confidence interval $\gamma$\% approximately $\gamma$\% of the time \citep{kuleshov2018accurate}.
For example, the model in the top panel of \fref{fig:calib} is not calibrated because only 20\% (2 out of 10) of the time the predictions are within the 90\% confidence interval, while the one in the bottom is calibrated.
Formally, according to \cite{kuleshov2018accurate}, for a given calibration dataset $D_\text{cal} = \{(x_i, y_i)\}_{i=1}^N$, a regression model is calibrated if
\begin{equation} \label{eq:calib:interval}
   \frac{\sum_{i=1}^N \mathbb{I} [y_i \leq F_i^{-1}(p) ] }{N}  \rightarrow p \ \text{for all}\ p \in [0, 1], 
\end{equation}
as $N\rightarrow\infty$.
Here, $F_i = P(Y<y_i)$ denotes the CDF of the random variable $Y$, and $F_i^{-1}$ is the corresponding quantile function (see \sref{sec:dist:fn}).
$\mathbb{I}[c]$ is the indicator function, which evaluates to 1 if the condition $c$ is true and 0 otherwise.
In other words, \eref{eq:calib:interval} means that, for a calibrated model, the empirically observed CDF from the data and the expected CDF by the model should match as the dataset size goes to infinity. 

\begin{figure}[bth!]
    \centering
    \includegraphics[width=.6\columnwidth]{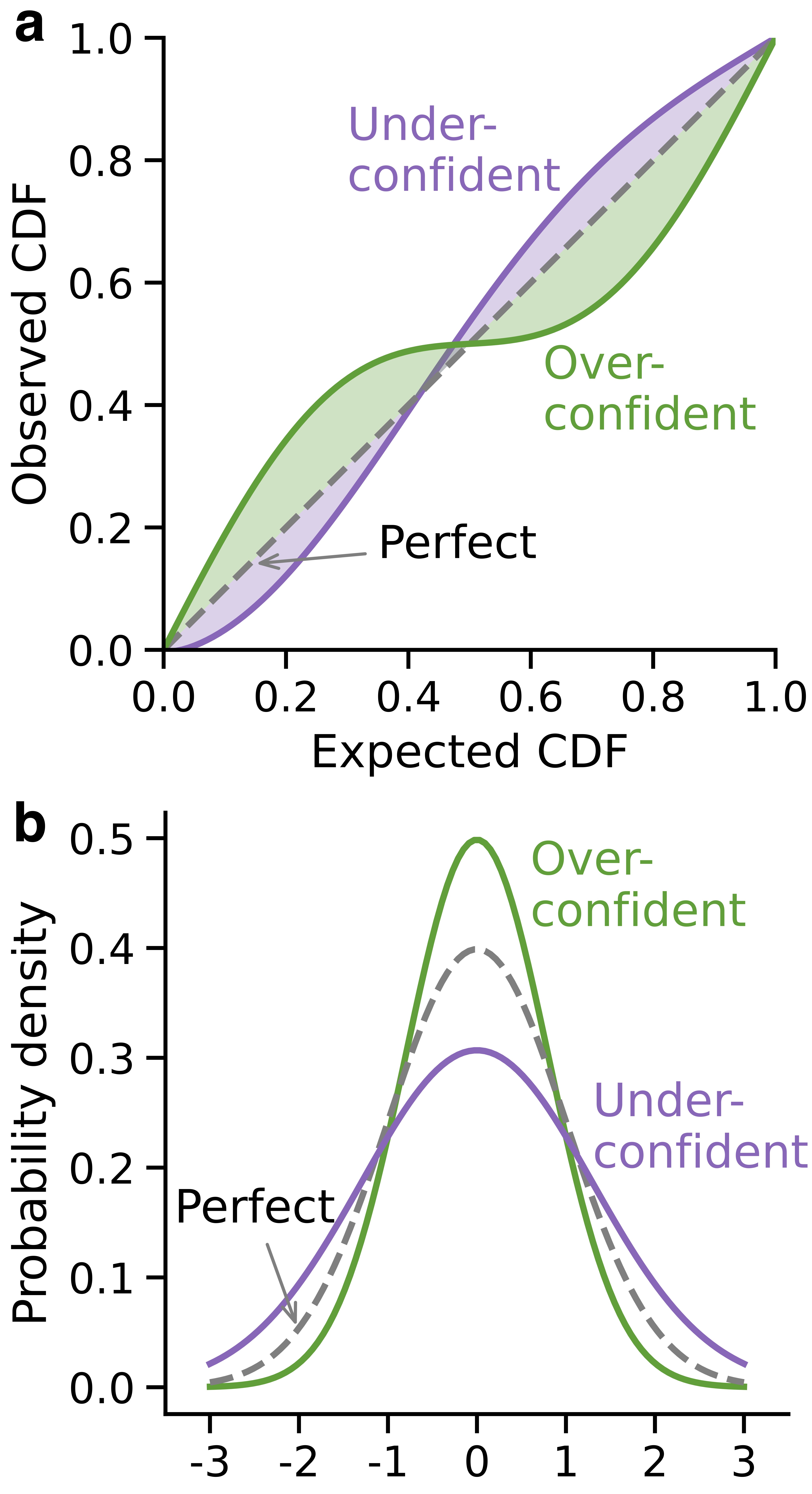}
    \caption{Calibration curves and probability density for the interval approach. 
    (a) Calibration curves for perfectly calibrated (diagonal grey), over-confident (green) and under-confident (purple) models.
    (b) The Gaussian probability densities correspond to the calibration curves in (a). 
    }
    \label{fig:calib:curve}
\end{figure}

In practice, \eref{eq:calib:interval} is evaluated for a selected number of $p$ values, and the \emph{calibration curve} is used to check the calibration level, which plots the observed CDF from the data versus the expected CDF by the model (\fref{fig:calib:curve}a).
With this, the calibration curve can be obtained as follows \citep{kuleshov2018accurate}:
\begin{enumerate}
    \item Discretize the expected CDF to a set of $M$ values $0 <p_1 < p_2 \dots < p_M < 1$. 
    The data is assumed to be generated from a Gaussian, $y \sim \mathcal{N}(\hat y, \sigma^2)$, where $\hat y$ and $\sigma^2$ are the predictions and the associated uncertainty, respectively \citep{kuleshov2018accurate, tran2022methods}.
    For example, if an ensemble method is used, then $\hat y$ and $\sigma^2$ are the ensemble mean and variance, respectively.
    For a Gaussian with mean $\hat y$ and variance $\sigma^2$, the expected CDF can be readily obtained (see \sref{sec:dist:fn}).
    We note that assuming a Gaussian distribution may not accurately reflect the true nature of the data in all cases.
    \item For each expected CDF $p_j$, compute the corresponding expected model output $y_j$.
    As mentioned above, the model output is assumed to follow a Gaussian; therefore, $y_j$ can be readily computed using the quantile function, $y_j = Q(p_j) = F^{-1}(p_j)$, discussed in \sref{sec:dist:fn}.
    \item For each expected CDF $p_j$, compute the corresponding observed CDF $\tilde p_j$. 
     With $y_j$ obtained in the previous step, $\tilde p_j$ is obtained as the empirical frequency $\tilde p_j = \vert \{ y_i \vert y_i \leq y_j, i = 1,2,\dots N \} \vert / N$ (left of \eref{eq:calib:interval}), where $\vert \cdot \vert$ denotes the size of a set, and $N$ is the total number of data points in the calibration dataset $D_\text{cal}$.
    In other words, $\tilde p_j$ is computed as the fraction of data points whose prediction $y_i$ is smaller than or equal to $y_j$.
    \item Create the calibration curve by plotting $(p_j, \tilde p_j)$ pairs for $j = 1,2,\dots M$.
\end{enumerate}

The calibration curve provides rich information.
First, for a perfectly calibrated model as defined in \eref{eq:calib:interval}, the calibration curve should be a diagonal line, meaning that the observed CDF from the data and the expected CDF by the model match with each other.
Therefore, a model's calibration could be qualified by the closeness of its calibration curve to the diagonal line.
In addition, the shape of the calibration curve could yield other insights into the predictive uncertainty of a model.
A calibration curve that is above the diagonal line at low expected CDF but below at high expected CDF suggests that the model is over-confident.
To understand this, let's focus on the low expected CDF region, e.g., at 0.2 in \fref{fig:calib:curve}a.
Here, the expected CDF is smaller than the observed CDF, meaning that the variance in the Gaussian distribution used to construct the model is smaller than the variance in the observed data (\fref{fig:calib:curve}b).
With a smaller expected variance (uncertainty), the model is over-confident.
On the other hand, an under-confident model has a calibration curve that is below the diagonal line at a small expected CDF but above a large expected CDF.

In \cite{kuleshov2018accurate} and \cite{levi2022evaluating}, the observed CDFs are interpreted as observed confidence intervals.
Because of the use of CDF, the interval here means $(-\infty, q_j]$.
This is different from the commonly used notion of confidence interval, which is typically specified as an interval around the mean.
For example, the 68\% confidence interval of a Gaussian distribution is $\mu \pm \sigma$.
Thus, to avoid confusion, we do not use ``confidence interval'' but directly use CDF as is also done in \cite{tran2022methods}.
Nevertheless, it is possible to interpret the calibration curve as the commonly used confidence interval.
Instead of CDF, one can employ the probability density function (PDF), considering symmetric intervals 
$\mu \pm \gamma_j$ of varying confidence level $0 <\gamma_1 < \gamma_2 \dots < \gamma_M < 1$ around the mean and examining the empirical frequency of the observed data belonging to each interval \citep{scalia2020evaluating}.

\subsubsection{Error based approach}

The error based approach directly compares the predicted uncertainty $\sigma^2$ and the expected square error between the model prediction $\hat y$ and the observed data $y$, stating that, for a calibrated model, the predicted uncertainty and the expected error should match \citep{levi2022evaluating}.
Formally, a regression model is calibrated if 
\begin{equation}\label{eq:calib:error}
   \mathbb{E}_{x,y} \left[(\hat y - y)^2 \cond \sigma^2 = u \right] \rightarrow u
\end{equation}
for any chosen positive $u$, 
where the expectation $\mathbb{E}$ is taken over the joint distribution of $x$ and $y$.
From the definition, no average over points with different values of $u$ is needed; thus, in principle, for each data point, one can correctly predict the expected error.
In practice, however, binning is performed to empirically evaluate \eref{eq:calib:error}.

\begin{figure}[bth!]
    \centering
    \includegraphics[width=0.6\columnwidth]{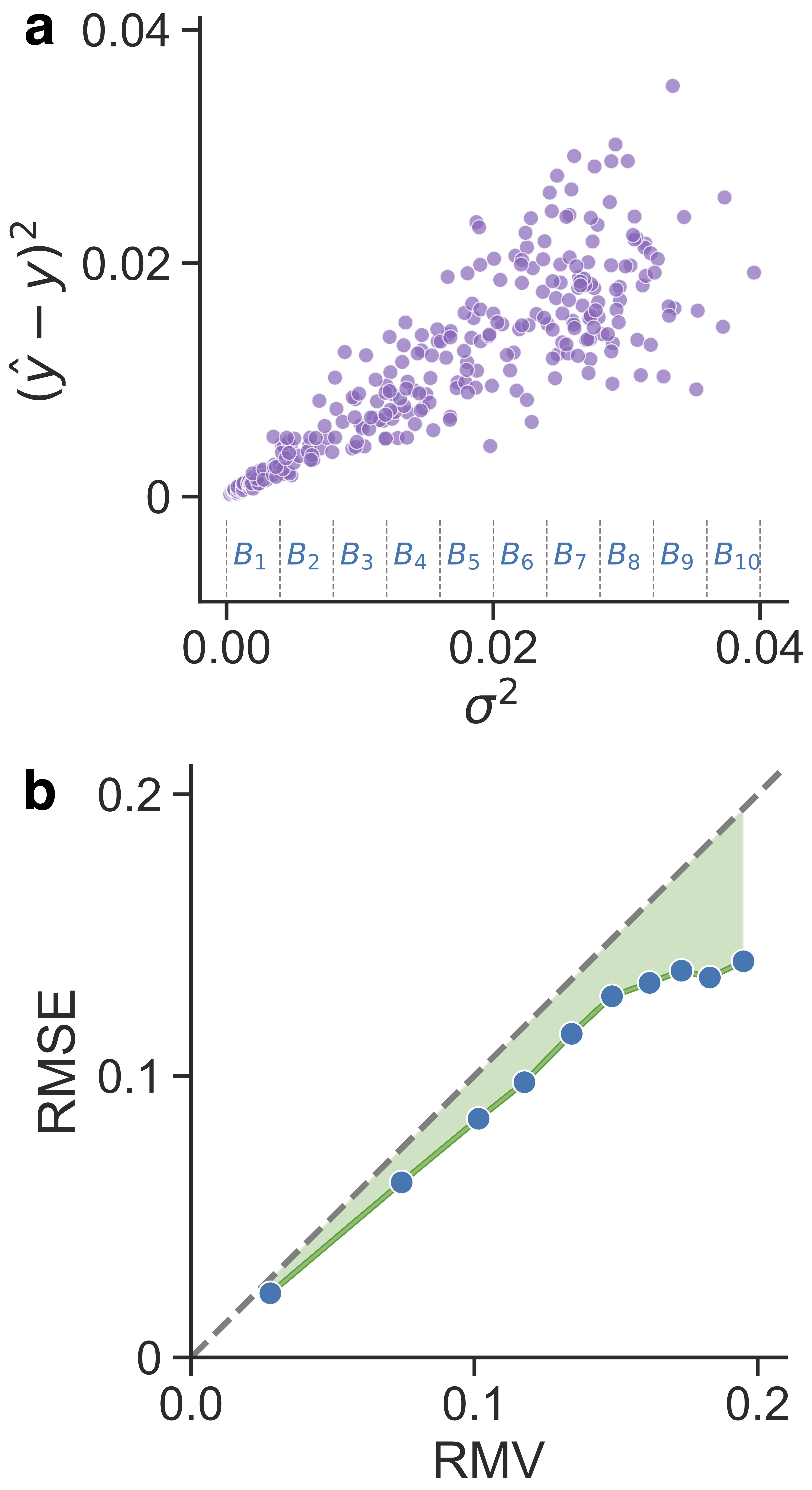}
    \caption{Reliability diagram for error based calibration approach. (a)
    Binning an example dataset of 400 data points into 10 equally separated intervals.
    (b) Calibration curve, where each blue dot represents the RMSE and RMV for each bin. 
    A perfectly calibrated model should follow the diagonal line.
    }
    \label{fig:reliability:diagram}
\end{figure}

Similar to the calibration curve in the interval based approach, here, a \emph{reliability diagram} can be created to diagnose the calibration level of a model, as follows \citep{levi2022evaluating}:
\begin{enumerate}
\item Sort the data points according to their predicted uncertainty $\sigma^2$, and then divide them into $M$ bins, $B_1, B_2, \dots B_M$.
For simplicity, the bin boundaries can be equally located from the minimum to the maximum of the uncertainty (\fref{fig:reliability:diagram}a).
\item For each bin $B_j$, calculate the root-mean variance (RMV),
$ \text{RMV}(j) = \sqrt{\frac{1}{|B_j|} \sum_{i \in B_j} \sigma^2_i},
$
and the root-mean-square error (RMSE), 
$ \text{RMSE}(j) = \sqrt{\frac{1}{|B_j|} \sum_{i \in B_j} ( \hat{y}_i -y_i)^2},
$
where $|B_j|$ is the number of data points in bin $j$.
\item Plot the $\text{RMSE}(j)$ against the $\text{RMV}(j)$ for each bin $j$. 
This plot is the reliability diagram
(\fref{fig:reliability:diagram}b).
\end{enumerate}

According to \eref{eq:calib:error}, if a model is perfectly calibrated, the RMV and RMSE should be equal for each bin;
therefore, the calibration curve should be a straight diagonal line in \fref{fig:reliability:diagram}b.
A larger deviation from the diagonal line suggests a more poorly calibrated model.
It is important to note that, unlike the interval based approach, here, the reliability diagram is not constrained to be within $[0, 1]$, but instead ranges between 0 and the maximum RMV or the maximum RMSE value.
Consequently, directly comparing the reliability diagrams of different models is not appropriate unless some form of normalization is performed.
Furthermore, the choice of the number of bins can have a significant impact on the results.
For instance, in \fref{fig:reliability:diagram}a, we chose 10 bins, and the last bin $B_{10}$ consists of only three data points, which is insufficient to obtain reliable statistics for RMV and RMSE.
One could consider using a smaller number of bins; however, if the number of bins is too small, the details might be averaged out.

\subsection{Metrics}
\label{sec:metrics}

Calibration plots offer a qualitative way to assess UQ methods.
For quantitative comparison, scoring metrics that can assign numerical scores to each UQ method become necessary.
Various metrics have been proposed, and we focus on the important ones that evaluate a UQ method from four different perspectives: calibration, precision, accuracy, and efficiency.

\subsubsection{Calibration error}
\label{sec:calib:error}

\textbf{Miscalibration area}.
For the interval based calibration approach, the closeness of a model's calibration curve to the perfect calibration curve (i.e., the diagonal line) can be quantified by calculating the area between them (e.g., the green area in  \fref{fig:calib:curve}a), called the \emph{miscalibration area} \citep{tran2022methods}.
A smaller miscalibration area indicates better calibration and a miscalibration area of 0 suggests an ideal calibration.

\textbf{Expected normalized calibration error (ENCE)}.
For the error based calibration approach, it does not make sense to calculate the area between a model's calibration curve and the perfect calibration curve and then compare across models, because the RMV and RMSE values are not bounded to be within $[0, 1]$ and different UQ methods can have varying ranges for RMV and RMSE.
To alleviate this, the ENCE can be used \citep{levi2022evaluating},
\begin{equation}
\text{ENCE} = \frac{1}{M} \sum_{j=1}^{M} \frac{ \vert \text{RMV}(j) -\text{RMSE}(j) \vert}{\text{RMV}(j)}, 
\end{equation}
where $M$ is the total number of bins used to generate the calibration curve.
Similar to the miscalibration area, a smaller ENCE indicates better calibration.

Miscalibration error and ENCE summarize the reliability by aggregating/averaging the errors between the predicted and perfect calibration curves, producing an overall assessment of a model's calibration.
One can also consider the maximum calibration difference between the curves to obtain the worst-case error. 
This becomes important in high-risk applications, e.g., in drug discovery and materials design for safety-critical systems.

\subsubsection{Precision}

Calibration is necessary but not sufficient for useful UQ analysis \citep{gneiting2007probabilistic}.
Recall that calibration only measures the statistical consistency between the predicted uncertainty and the observed data, but does not provide information about the distribution of the predictions themselves.
This aspect is related to the \emph{precision} of the predictions, and metrics such as \emph{sharpness} and \emph{dispersion} have been proposed to quantify it \citep{kuleshov2018accurate, levi2022evaluating}.
However, according to \cite{gneiting2007probabilistic}, these metrics should be considered secondary after calibration.
A major reason for this is that they are properties of the predicted uncertainty alone and do not capture the relationships between uncertainty and accuracy.

\textbf{Sharpness}. 
A well-calibrated model with more precise predictions (small uncertainty estimates) would be more informative and useful than a less precise model (large uncertainty estimates) \citep{gneiting2007probabilistic}.
This idea can be quantified by the \emph{sharpness}, defined as \citep{kuleshov2018accurate, tran2022methods}:
\begin{equation}
    \text{SHA} = \sqrt{\frac{1}{N} \sum_{i=1}^N \text{Var}(F_i)},
\end{equation}
where $\text{Var}(F_i)$ is the variance of the random variable whose CDF is $F$ for data point $i$.
In practice, this can be evaluated as $\text{SHA} = \sqrt{\frac{1}{N} \sum_{i=1}^N \sigma_i^2}$, where $\sigma_i^2$ is the predicted variance (uncertainty) for data point $i$.
The more precise the predictions, the sharper the model (smaller SHA value), and the sharper the better.

\textbf{Dispersion}.
Another dimension involves the dispersion of the uncertainty.
One can obtain perfectly calibrated uncertainty if a model always outputs the same constant uncertainty which matches the empirical frequency across the entire distribution \citep{scalia2020evaluating, tran2022methods}. 
Such an uncertainty estimate is not informative or useful because it remains unchanged regardless of the input data provided to the model.
\cite{levi2022evaluating} propose the \emph{coefficient of variation} to measure the dispersion of the uncertainty estimates,
\begin{equation}
   C_v = \frac{1}{\mu_\sigma}\sqrt{
   \frac{1} {N-1}
   \sum_{i=1}^N (\sigma_i - \mu_\sigma)^2 }, 
\end{equation}
where $\sigma_i$ is the predicted standard deviation, $\mu_\sigma$ is the mean of the standard deviations and $N$ is the total number of data points.
A $C_v$ value of 0 means the same constant uncertainty for all data points, not a useful uncertainty estimate. 
Higher $C_v$ is preferred so that the uncertainty for different data points can be distinguished.

\subsubsection{Accuracy}

The introduction of a UQ method to a model can affect the model's prediction accuracy.  
For example, it has been observed that graph NNs for chemical property prediction trained with ensemble and MC dropout methods yield higher accuracy when compared with the same model trained using maximum likelihood \citep{scalia2020evaluating}.
But this may not always be the case.
So, it is crucial to examine prediction accuracy as well.
The two most widely used accuracy metrics are mean absolute error (MAE):
\begin{equation}
\text{MAE} = \frac{1}{N} \sum_{i=1}^{N} |\hat y_i - y_i|, 
\end{equation}
and root-mean-square error (RMSE):
\begin{equation}
\text{RMSE} = \sqrt{\frac{1}{N} \sum_{i=1}^{N} (\hat y_i - y_i)^2},
\end{equation}
where $\hat{y}_i$ is model prediction, $y_i$ is the corresponding reference value, and $N$ is the number of data points.
MAE measures the average absolute difference between the predicted and reference values, treating all errors equally. 
On the other hand, by squaring the errors before averaging, RMSE gives higher weight to larger errors, making it more sensitive to outliers.
Lower values of MAE and RMSE indicate better agreement between predictions and reference data.

\subsubsection{Efficiency}

Computational tractability and efficiency are essential for a UQ method to be practically usable. 
Even if a method is highly calibrated, precise, and accurate, it may not be suitable for real-world applications if it is computationally too demanding in terms of both time and memory. 
Unfortunately, efficiency is often ignored in existing studies of UQ methods for atomistic ML.

\textbf{Training and inference time}.
A straightforward way to compare time efficiency is by tracking the total runtime of a model to obtain the prediction and uncertainty, which is usually the main concern for practical atomistic ML applications.
However, a UQ method's total runtime is highly dependent on the underlying model's speed.
To focus on the UQ method itself, we analyze training and inference efficiency separately: the number of models required to be trained and the number of model executions needed to obtain the uncertainty at inference.
The ensemble approach is not efficient because multiple models need to be trained, and multiple model executions must be carried out to get the uncertainty at inference.
Approaches such as MVE and evidential regression are on the opposite end of the spectrum: a single model at training and a single execution at inference.
Methods like MC dropout lie in between these two extremes.
The training/inference efficiency for all UQ methods discussed in \sref{sec:uq} is listed in \tref{tab:summary}.

\textbf{Memory}. 
In addition to time efficiency, memory efficiency should be another consideration when evaluating the UQ methods. 
This is, again, largely dependent on the underlying model to which a UQ method is applied.

\subsubsection{Other metrics}

Besides the above-mentioned ones, other metrics have also been used to evaluate UQ performance, particularly, in assessing a model's calibration.
These include ranking correlation and NLL, among others, which can be used together with the calibration metrics discussed in \sref{sec:calib:error}.

\textbf{Ranking correlation}.
For a UQ method, we expect that a high uncertainty suggests a large prediction error.
So, there should be a monotonic relationship between the uncertainty $\delta$ and the prediction error $\epsilon$ for a well-calibrated model \citep{tan2023single, varivoda2023materials}.
This can be quantified with Spearman's rank correlation coefficient.
For a set of uncertainties and errors $\{(\delta_i, \epsilon_i)\}_{i=1}^{N}$, we first obtain ranked sequences of the uncertainties $R_\delta$ and the errors $R_\epsilon$, separately, and then compute the Spearman's rank correlation coefficient as 
\begin{equation}
   r_s  
   = \frac{\text{Cov}(R_\delta, R_\epsilon)}{\sigma_{R_\delta} \sigma_{R_\epsilon}}, 
\end{equation}
where $\text{Cov}$ denotes the covariance between two variables and $\sigma$ denote the standard deviation of a variable.
The values of $r_s$ are within the range of $[-1, 1]$, with $-1$ or $1$ suggesting a perfect monotonic relationship between the uncertainty and the error, and 0 being the worst case, indicating that there is no correlation.

\textbf{NLL}.
NLL is a standard measure of a model's fit to the data which combines both the accuracy and the uncertainty in one measure. 
With a set of predictions and the associated uncertainties (i.e., variance $\sigma^2$),
\eref{eq:NLL} can be directly used to obtain the NLL.
Despite its popularity, NLL has been criticized for the lack of robustness \citep{gneiting2007probabilistic}.
It is hypersensitive to small changes and is unbounded, with acceptable values ranging from $-\infty$ to $+\infty$ 
\citep{gneiting2007strictly, selten1998axiomatic}.

\subsection{Benchmark Studies}
\label{sec:eg}

\begin{figure*}[tbh!]
    \centering
    \includegraphics[width=2\columnwidth]{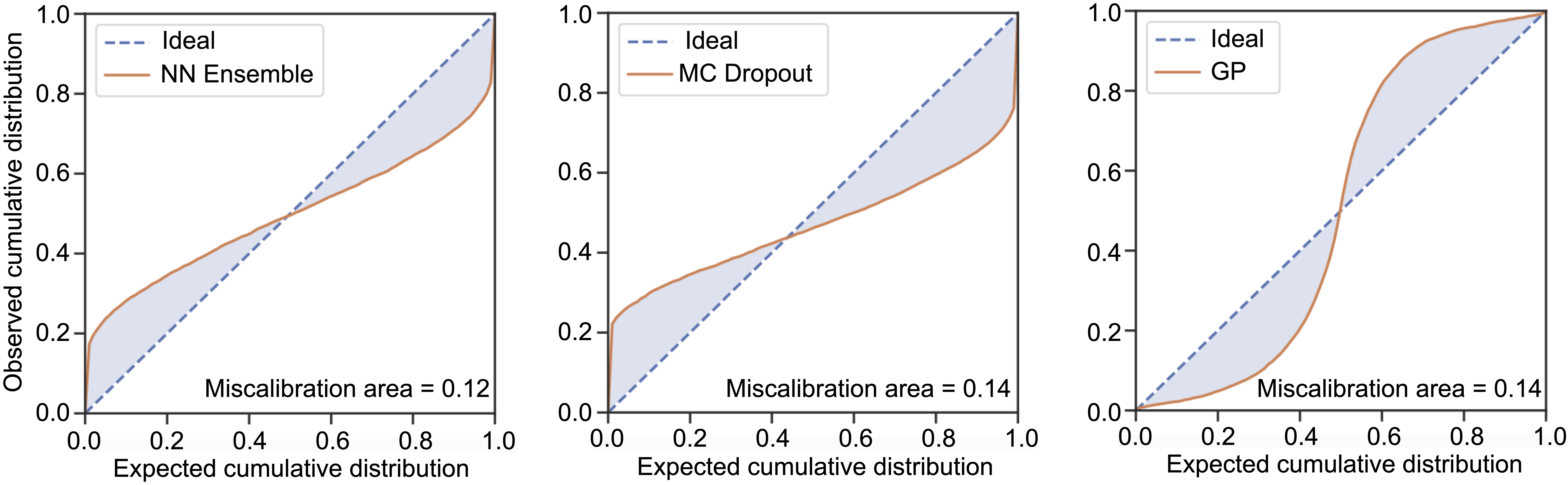}
    \caption{Comparison of calibration curves displaying miscalibration area of various UQ methods. 
    NN ensemble, here, refers to NNs trained with random parameter initializations.  
    GP: Gaussian process, a probabilistic modeling approach that inherently models predictive uncertainty \citep{rasmussen2003gaussian}. Images adapted from \cite{tran2022methods} under a CC BY 4.0 DEED license.}
    \label{fig:calib:example}
\end{figure*}

\begin{figure*}[tbh!]
    \centering
    \includegraphics[width=2\columnwidth]{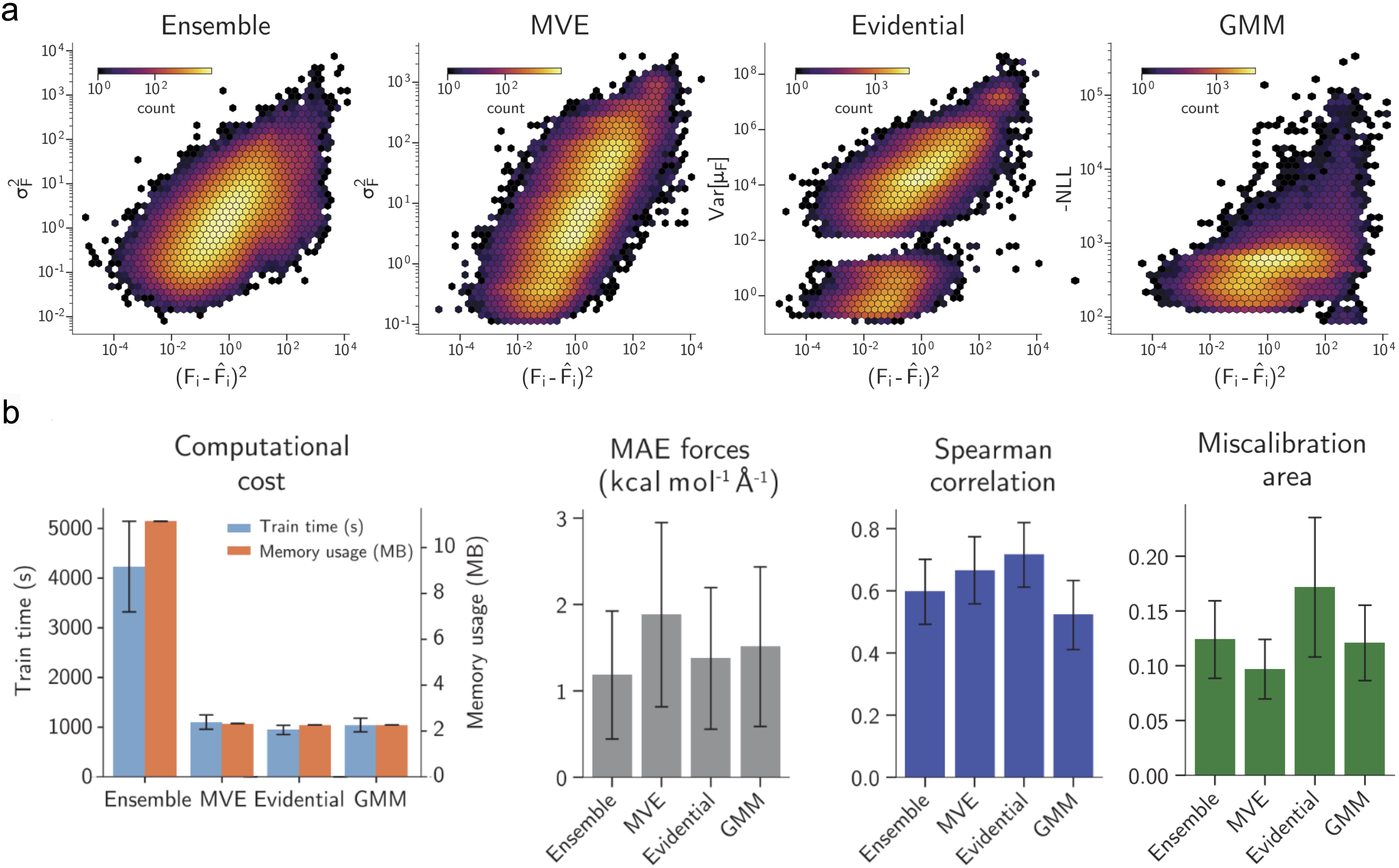}
    \caption{Comparison of various UQ methods on the rMD17 dataset. 
    (a) Predicted uncertainties as a function of squared errors of atomic forces and 
    (b) computational cost, MAE of atomic forces, Spearman correlation, and miscalibration area of various UQ methods.
    Ensemble, here, refers to NNs trained with random parameter initialization. 
    Images adapted from \cite{tan2023single} under a CC BY 4.0 license. }
    \label{fig:metrics:example2}
\end{figure*}

\begin{figure}[tbh!]
    \centering
    \includegraphics[width=1.0\columnwidth]{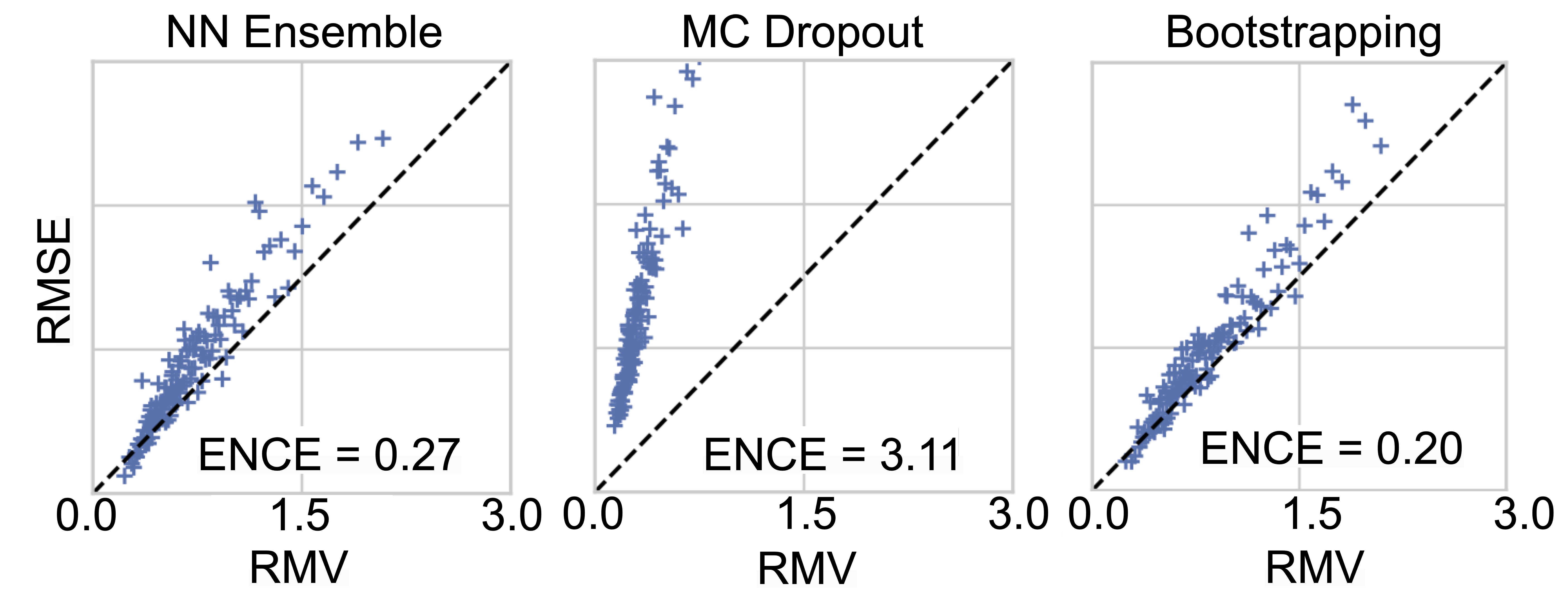}
    \caption{Comparison of error based calibration curves of three UQ methods on the QM9 dataset. 
    NN ensemble, here, refers to NNs trained with random parameter initializations.  
    Image adapted from \cite{scalia2020evaluating}. Copyright 2020, American Chemical Society.}
    \label{fig:metrics:example1}
\end{figure}

Using metrics as those discussed in \sref{sec:metrics}, several benchmark studies have attempted to evaluate the performance of various UQ methods on diverse atomistic ML datasets 
\citep{scalia2020evaluating, hirschfeld2020uncertainty, tran2022methods, hu2022robust, tan2023single, varivoda2023materials}.
A major goal is to identify UQ methods that can perform well across metrics and datasets, thus providing practical guidance for selecting appropriate ones for chemical and materials applications. 
Several general observations can be made from these benchmark studies.

First, the UQ methods perform differently on different metrics; a method that works well on one metric can fall short on another.
For example, \cite{tran2022methods} have trained various ML models to predict the adsorption energies of small molecules on metal surfaces calculated from DFT.
Although all trained ML models reported in \cite{tran2022methods} have similar accuracy (MAE of $\sim$0.20~eV) and miscalibration area ($\sim$0.13, \fref{fig:calib:example}), their performance on precision varies a lot.
For example, MC dropout is much sharper 
than NN ensemble (SHA: 0.09 versus 0.14), but has a lower dispersion ($C_v$: 0.82 versus 1.06) (see section 3, \cite{tran2022methods}, for more details). 
Even for UQ methods that have, for example, a similar miscalibration area as illustrated in \fref{fig:calib:example}, the shape of the calibration curves can be drastically different, indicating different modes of miscalibration. 
\cite{tan2023single} observed similar behavior using the rMD17 dataset \citep{christensen2020role} of energies of small molecules.

Second, performance varies by dataset; for a given metric, different UQ methods can have varying error levels across datasets.
For example, \cite{tan2023single} observed that, for the rMD17 dataset of energies of small molecules, NN ensemble exhibits smaller miscalibration error than the evidential regression method (\fref{fig:metrics:example2}b).
\cite{varivoda2023materials} have found similar behaviors for the dataset of formation energies of crystals \citep{jain2013commentary} and the dataset of surface adsorption energies of metal alloys \citep{mamun2019high}.
However, they found that, for the dataset of band gaps of MOFs \citep{rosen2021machine}, the miscalibration errors are the same for the ensemble and evidential regression methods.

Third, ensemble methods appear to be a reliable UQ approach in general. 
For example, \cite{tan2023single} compared the ensemble method versus the MVE, evidential regression, and GMM deterministic methods for both in-domain (rMD17 and ammonia) and out-of-domain (silica glass) tasks.
Using metrics such as MAE, Spearman correlation, and miscalibration area (\fref{fig:metrics:example2}), they have found that single-deterministic methods struggle to consistently perform better across each in-domain and out-of-domain task and that the ensemble method still remains the most reliable choice. 
For ensemble methods, different ways to generate the ensemble can result in different performances.
For example, \cite{scalia2020evaluating} have compared three ensemble methods---NN ensemble with different parameter initialization, bootstrapping, and MC dropout---using the miscalibration area, ENCE, sharpness, and dispersion metrics.
(Recall from \ref{sec:dropout} that MC dropout can be regarded as an ensemble method in practice.)
Evaluated on various MoleculeNet benchmarking datasets \citep{wu2018moleculenet}, their finds indicate that NN ensemble with different parameter initialization and bootstrapping consistently outperform MC dropout (\fref{fig:metrics:example1}).

Despite the robustness of the ensemble approach, there are contradictory studies showing that ensemble methods may not always be the most reliable choice.  
For example, \cite{hirschfeld2020uncertainty} have demonstrated that the stacking methods that sequentially combine multiple weak models to produce the uncertainty estimate can be more consistent than the ensemble approach. 
\cite{heid2024spatially} have shown that the uncertainty of a single prediction obtained from an ensemble approach cannot be directly correlated with the absolute error per atom. 
This is because the absolute error is distributed along a normal distribution, with its width determined by the uncertainty arising from model variance. 
To address this, they developed an approach that uses locally aggregated uncertainties to identify high-error local substructures, enabling the resolution of absolute errors on an atomic scale.

Fourth, off-the-shelf metrics may not be directly applicable. 
The metrics, particularly those related to calibration discussed in Section \ref{sec:metrics}, have been primarily developed within the ML community using non-chemical datasets. 
Their suitability for chemical and materials problems is not guaranteed. 
For instance, the error based calibration approach conventionally performs binning directly on the predicted uncertainty, represented by the variance $\sigma^2$ in \fref{fig:reliability:diagram}a. 
Chemical and materials data can have numerous small variance values. Therefore, it might be more appropriate to conduct binning after transforming the variances using a logarithmic function to avoid cluttering the bins at small variance values, as demonstrated in \fref{fig:metrics:example2}a.

Benchmark studies so far have been valuable in highlighting how different UQ methods can have varied performances depending on the choice of metrics and dataset.
However, these studies consistently indicate that the question of which UQ method to select in practice remains unresolved.
This appears to be discouraging.
Nevertheless, there are guiding principles based on ease of use, efficiency, and uncertainty propagation.
Ensemble methods provide a strong baseline and are straightforward to implement, making them a go-to choice for many applications. 
However, single-pass models (e.g., MVE, evidential, and GMM) are generally more efficient than ensemble methods (Figure \ref{fig:metrics:example2}b), so they can be good choices for resource-bounded applications.
Another consideration is UP; the chosen propagation method may make certain UQ methods more suitable. 
This will be further discussed in \sref{sec:up}.

\subsection{Recalibration}
\label{sec:recalibration}

So far, we have discussed ways for evaluating UQ methods in terms of their calibration, precision, accuracy, and efficiency, and we have provided some examples. 
But if a model's performance is unsatisfactory, are there any approaches to improve it? 
The answer is yes, and, here, we will concentrate on calibration.

Informally, the calibration problem can be described as follows: given a trained model $U$ that can generate predictive uncertainty $\delta = U(x)$, we train another recalibration model $R$ such that the output of the composed functions $\hat\delta = R(U(x))$ is calibrated. 
The model $R$ should be trained on a separate recalibration dataset $D_\text{cal}$ distinct from the datasets used for model parameter optimization or hyperparameter tuning.
Below, we discuss two recalibration methods, both of which are applicable to the interval based calibration approach introduced in \sref{sec:calib:interval}.

\textbf{Variance scaling}. 
For the interval based calibration, the model prediction, $y$, is set to take the form of a Gaussian $y \sim \mathcal{N}(\mu, \sigma^2)$.
The model can be under-confident or over-confident depending on the scale of the variance $\sigma^2$. 
To recalibrate it, that is, making the calibration curve move toward the diagonal line in \fref{fig:calib:curve}a, we can train a linear model $R(\sigma^2): \hat \sigma^2 = a\sigma^2 + b $ to scale the variance. 
The parameters $a$ and $b$ can be determined by, e.g., minimizing a calibration NLL loss \citep{tan2023single},
$\frac{1}{2} \sum_{i=1}^{N} \left(\log[2\pi (a\sigma^2 + b )] + {(y_i - \hat y_i)^2}/{(a\sigma^2 + b)} \right),
$
using a recalibration dataset $D_\text{cal}$ consisting of $N$ data points.
Once the optimal $a$ and $b$ are obtained, the new variance $\hat \sigma^2$ for the Gaussian is known for every data point, which can then be used to regenerate the calibration plots.

\textbf{Isotonic regression}.
The variance scaling approach still assumes that the model prediction follows a Gaussian distribution. 
The true observed data distribution, however, may not be Gaussian. \cite{kuleshov2018accurate} proposed a recalibration approach based on isotonic regression, which is effective even for non-Gaussian cases.
Given a recalibration dataset $D_\text{cal}$, this approach begins by transforming it into a processed recalibration dataset $\hat D_\text{cal} = \{(q_j, \tilde q_j)\}_{j=1}^M$, where $q_j$ and $\tilde q_j$ are obtained using the same procedures described in \sref{sec:calib:interval}. 
Using $\hat D_\text{cal}$, an isotonic regression model $\tilde q_j = R(q_j)$ is then trained.
Isotonic regression is a technique for fitting a free-form line to map a sequence of inputs ($q_j$ in this case) to a sequence of observations ($\tilde q_j$ in this case) such that the fitted line is non-decreasing everywhere and lies as close to the observations as possible \citep{fielding1974statistical}. 
Isotonic regression is chosen because it accounts for the fact that the true calibration curve is monotonically increasing.
Once the isotonic regression model is trained, new uncertainty $\hat \delta$ can be generated, and new calibration plot can be produced.

\section{Uncertainty propagation}
\label{sec:up}

For most chemical and materials problems, quantifying a model's predictive uncertainty is not sufficient; 
often, we are also interested in understanding how the uncertainty propagates to a physical QoI that can be obtained from physics-based modeling using the model.
For example, if the uncertainties in energy and forces are known for an interatomic potential, and MD simulations are employed to compute a material property like thermal conductivity, we would naturally hope to know the uncertainty in the calculated thermal conductivity.
Similarly, if microkinetic modeling is used to investigate chemical reaction dynamics, we need to determine how the uncertainty in reaction rates propagates to and affects the concentrations of the chemical species.

We define the uncertainty propagation (UP) problem as follows:
Given a model $y = f(x; \theta)$ that can provide predictive uncertainty, we aim to determine the uncertainty in a QoI $z$ that is a function of the model output, i.e., $z = g(y) = (g \circ f)(x; \theta)= h(x; \theta)$, where $h = g\circ f$ is defined as the composition of $g$ and $f$.
In other words, we investigate how the uncertainty in the model parameters $\theta$ and the training data propagate to the QoI $z$.
Typically, $g$ is not an ML model but rather a physics-based simulation technique, such as MD or microkinetic modeling, as mentioned above.

While UQ has been reasonably well investigated in atomistic ML for chemical and materials applications, UP remains a relatively unexplored area. 
Nevertheless, UP is essential for building confidence in the results. It provides a comprehensive assessment of the entire modeling pipeline, enabling the evaluation of the robustness of the final results. 
Furthermore, UP helps identify the most influential uncertainty sources, guiding targeted efforts to refine and improve the modeling pipeline.
In this section, we introduce some of the efforts in UP for MD and microkinetic simulations.

\subsection{Bayesian propagation}

Given the posterior over model parameters $p(\theta\cond D)$ (see \sref{sec:bayesian}), the distribution of a QoI $z$ can be written as 
\begin{equation} \label{eq:qoi:bayesian}
   p(z\cond x, D)  = \int p(z\cond x, \theta) p(\theta \cond D)\,\text{d}\theta , 
\end{equation}
where $p(z\cond x, \theta)$ is the likelihood of $\theta$ to observe $z$ given the composed model $z = h(x; \theta)$. 
The integration, generally, cannot be analytically evaluated for most QoI in MD and microkinetic simulations.    
To address this, again, sampling techniques can be employed.
\eref{eq:qoi:bayesian} can be approximated by \citep{angelikopoulos2012bayesian} 
\begin{equation} \label{eq:qoi:sample}
   p(z\cond x, D)  = \frac{1}{M} \sum_{i=1}^M p(z\cond x, \theta_i).
\end{equation}
Assuming the likelihood $p(z\cond x,\theta)$ is a Gaussian with $h(x; \theta)$ as its mean, then the predictive mean and variance of \eref{eq:qoi:sample} can be respectively expressed as 
\begin{equation}
   \bar z 
   = \frac{1}{M} \sum_{i=1}^M z_i
    = \frac{1}{M} \sum_{i=1}^M h(x; \theta_i) ,
\end{equation}
and
\begin{equation}
   \sigma^2_z = \frac{1}{M-1} \sum_{i=1}^M (z_i - \bar z)^2 .
\end{equation}

The sampling-based Bayesian UP means selecting multiple sets of model parameters $\theta$, computing the QoI $z_i$ using each parameter set, and then calculating their mean as the final prediction and their variance as the uncertainty.  
The sampling of the parameters can be done via MC methods, such as MCMC \citep{berg2004markov} and transitional MCMC \citep{ching2007transitional}.

Practically, it can also be viewed as an ensemble approach, where each realization of the parameters is a member of the ensemble.
Therefore, for the UQ methods discussed in \sref{sec:uq}, this UP method can be directly applied to models trained with MC dropout and the ensemble approach, but not for the others.

\begin{figure}[tbh!]
    \centering
    \includegraphics[width=.8\columnwidth]{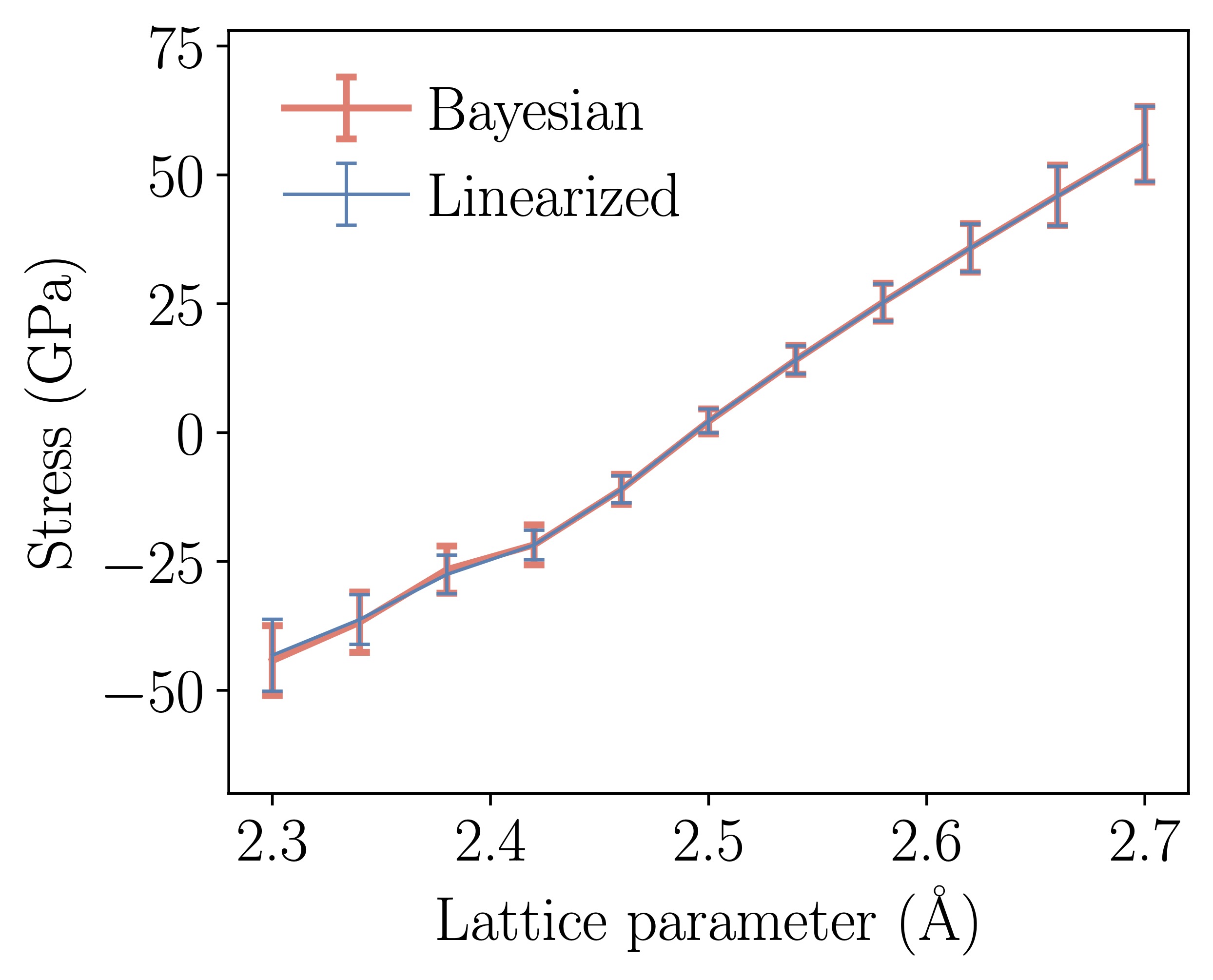}
    \caption{Uncertainty propagation in molecular dynamics computation of mechanical stresses in a monolayer graphene.
    Error bars indicate uncertainty level.
    Adapted from \cite{wen2020uncertainty} with a CC BY 4.0 license. 
}
    \label{fig:up:stress}
\end{figure}

Using an NN interatomic potential trained with MC dropout in MD simulations, \cite{wen2020uncertainty} adopted this sampling-based approach to propagate the uncertainty in atomic forces to the mechanical stress on monolayer graphene (\fref{fig:up:stress}).
As expected, the uncertainty in stress increases as the graphene layer is compressed or stretched from its equilibrium lattice parameter of 2.466~\AA. 

Uncertainty and error propagation in microkinetic modeling have long been studied.
Notably, uncertainties as a result of different DFT functionals such as those from the BEEF family (Bayesian error estimation functional) \cite{mortensen2005bayesian} have been extensively investigated.
In BEEF, an ensemble of exchange-correlation functionals is created, and the uncertainty in the ensemble can be propagated to downstream microkinetic modeling.
For example, uncertainty from BEEF-vdW \citep{wellendorff2012density} has been used in  computational catalysis to assess the reliability of
calculated reaction rates \citep{lu2022quantifying}, 
identified reaction mechanisms \citep{kreitz2023automated}, 
and coverage analysis \citep{wang2019propagating},
among others.
Recently, various studies have investigated the effects of uncertainty from ML models on microkinetic modeling.
\cite{li2023bayesian} was able to propagate the uncertainty in stoichiometric coefficients and parameters in the Arrhenius law \citep{arrhenius1889dissociationswarme, arrhenius1889reaktionsgeschwindigkeit} to the concentration of chemical species in biodiesel production reaction systems, using the so-called Bayesian chemical reaction NNs. 
In a microkinetic study of ethanol steam reforming reactions, \cite{xu2023microkinetic} investigated the propagation of errors in binding energy predicted by ML models to kinetic properties such as reaction rates. 
Their results demonstrated that the preferred reaction pathway varies depending on the used ML model.

\subsection{Linearized propagation}

Linearized UP is a general approach that can be used together with all UQ methods discussed in \sref{sec:uq}.
Let $\delta_y$ be the uncertainty associated with the output $y$ of an ML model $y = f(x; \theta)$.
Then for a QoI $z$ that is a function of the model output, $z = g(y)$, we do a first-order Taylor expansion at $y_0$ to obtain $ z = g(y_0) + \frac{\partial g}{\partial y}   (y - y_0)$.
With this linearization, the uncertainty in $z$ can be expressed as \citep{arras1998introduction}:
\begin{equation} \label{eq:uq:linear1}
    \delta_z  =  \frac{\partial g}{\partial y}  \delta_y,
\end{equation}
meaning that the uncertainty in $y$ can be propagated to $z$ by multiplying the gradient. 
In general, if $g$ is a function of multiple independent inputs, i.e., $z = g(y_1, y_2, \dots, y_M)$, the uncertainty in $z$ can be written as \citep{arras1998introduction}: 
\begin{equation} \label{eq:uq:linear2}
    \delta_z^2  = \sum_i^M \left(\frac{\partial g}{\partial y_i}\right)^2 \delta_{y_i}^2 .
\end{equation}

Once the uncertainty is obtained from a UQ method, it can be readily propagated to the QoI $z$, using  \eref{eq:uq:linear1} or \eref{eq:uq:linear2}.
However, there are two challenges in applying this approach in practice.
First, it may not be immediately clear how to get the gradients of $g$, which represents physics-based simulation techniques such as MD and microkinetic modeling.
Second, if the function $g$ is linear, this approach is exact; but for nonlinear function, the linearization can introduce large errors \citep{cho2015improvement}.

The linearized UP approach becomes very attractive if the challenges are overcome.
For example, \cite{wen2020uncertainty} reformulated the integration algorithm in an MD simulation, and managed to propagate the uncertainty in atomic forces to stresses.
As seen from \fref{fig:up:stress}, the predicted mean stress and uncertainty agree very well with those from the sampling-based Bayesian approach.
The linearized approach is computationally more efficient than the Bayesian approach, since it only needs one model evaluation, while the Bayesian approach requires multiple model evaluations.

\subsection{Sensitivity analysis}

Sensitivity analysis examines how the uncertainty in a model's output can be apportioned to various sources of uncertainty in its inputs. 
Ideally, uncertainty and sensitivity analyses should be conducted in tandem; by working together, they can pinpoint the most critical input parameters, offering crucial insights into the model's reliability and providing guidance for further model refinement.
Although these techniques have not been widely employed together with ML models yet, we expect this to happen soon in the near future. 
To illustrate their potential, we introduce some of their usage with classical non-ML models.

Given a QoI $z = g(y)$, we can perturb the input by some amount $\Delta y$ and observe the change in the output $\Delta z$. 
This change, $\Delta z$, can be interpreted as the propagated uncertainty if we set the uncertainty in $y$ as $\Delta y$.
Typically, the input $y$ lies in an $M$-dimensional parameter space, i.e., $y = [y_1, y_2, \dots, y_M]$; then, depending on how $\Delta y$ is chosen, we get \emph{local sensitivity} and \emph{global sensitivity}.
Local sensitivity refers to perturbing each individual parameter $y_i$ and observing its effects on $z$ separately.
Global sensitivity refers to exploring the entire parameter space simultaneously, considering interactions between the parameters.

Sensitivity analysis is an integral part of microkinetic modeling \citep{motagamwala2020microkinetic}.  
Local sensitivity analysis like the derivative-based technique \citep{dopking2018addressing} and global sensitivity analysis such as 
the Sobol' method \citep{sobol2001global} and the Morris method \citep{morris1991factorial} have been widely employed.
For example, \cite{bensberg2023uncertainty} have recently leveraged these techniques to automatically refine structures, reaction paths, and energies in chemical reaction networks, and successfully identified a small number of elementary reactions and compounds that are essential for reliably describing the kinetics of 
the Eschenmoser--Claisen rearrangement reactions \citep{wick1964claisen} of allyl alcohol and of furfuryl alcohol.
In addition, their Morris sensitivity analysis also provides the uncertainty in the predicted concentrations.
Similarly, \cite{kreitz2021quantifying} applied global uncertainty assessment and sensitivity analysis to explore parametric uncertainties in microkinetic models for CO$_2$ hydrogenation on the (111) surface of Ni. 
By systematically generating numerous mechanisms, they demonstrated how uncertainty quantification can identify feasible models and optimize predictions within the uncertainty space.

Sensitivity analysis has also been applied to quantify the uncertainty in QoI obtained from MD simulations.
Information-theoretic approaches provide a powerful framework for this purpose \citep{kurniawan2022bayesian}.
For a QoI $z$ that can be obtained from an MD simulation, an upper bound for the uncertainty in $z$ can be obtained as  
\citep{pantazis2013relative, tsourtis2015parametric, dupuis2016path}:
\begin{equation}\label{eq:fisher:bound}
\vert \mathbb{E}_{\theta+  \Delta\theta}[z] - \mathbb{E}_{\theta}[z] \vert
\leq 
\sqrt{\text{Var}_{{\theta}}[z]}  
\sqrt{ \Delta\theta \mathcal{I}(\theta)  \Delta\theta}
\end{equation}
upon parameter perturbation $\Delta\theta$, where Var denotes the variance, and $\mathcal{I}(\theta)$ is the Fisher information, which measures the amount of information that the trained model carries about its parameter $\theta$ \citep{cover2012elements, wen2019thesis}.
This provides an efficient way to investigate the reliability of MD simulations.
Using this approach, \cite{wen2017force} studied the thickness of an MoS$_2$ sheet and found that \eref{eq:fisher:bound} provides a tight bound, demonstrating high reliability of the MD predictions.
Although Fisher information can provide useful insights into the uncertainty in MD simulations, the overall analysis is restricted to the perturbations only in the vicinity of the equilibrium model parameters.

\section{Summary and Outlook}

In this work, we have provided a comprehensive overview of the UQ approaches for atomistic ML.
The UQ methods are classified into three main categories: probabilistic, ensemble, and feature space distance. 
The similarities, differences, and connections between them were discussed to provide an overall overview of the methods.
We have discussed metrics to evaluate the performance of these UQ methods from different angles, focusing on calibration, precision, accuracy, and efficiency.
In addition, we have emphasized the importance of UP in downstream chemical and materials applications of the ML models.

We deliberately exclude some important but advanced topics to make the presentation more accessible and avoid further complications. 
For example, we chose to focus on UQ for NNs and ignore other methods such as Gaussian processes \citep{rasmussen2003gaussian}, which inherently provide predictive uncertainty. 
For the use of Gaussian processes in materials and molecular problems, we refer readers to the thorough review by \cite{deringer2021gaussian}.
Additionally, we do not explicitly discuss whether the aleatoric or epistemic uncertainty is modeled by a UQ method; 
instead, we provide the total uncertainty, as it is the combined effect of both sources that is relevant for most practical purposes.
Nevertheless, works such as \cite{gustafsson2020evaluating}, \cite{gawlikowski2023survey}, and \cite{heid2023characterizing} provide further discussion on this topic.

We have identified several challenges in UQ and UP for atomistic ML, along with potential opportunities to address these challenges. 
First, existing benchmark studies suggest that the performance of UQ methods is highly dependent on the datasets and metrics being used. There is no universal UQ method that consistently outperforms others in all scenarios. Thus, there is a high demand for a set of best practices and guidelines for UQ in atomistic ML. These guidelines should provide recommendations on choosing appropriate UQ methods based on factors such as the nature of the dataset, the complexity of the ML model, the available computational resources, and the downstream applications of the ML model.

A second challenge is the large miscalibration of existing UQ methods. 
The calibration curves of many UQ methods can deviate significantly from the diagonal line, suggesting that the predicted uncertainties do not match well with the observed errors. 
A straightforward solution is to perform uncertainty recalibration. 
Uncertainty recalibration for ML models is a relatively new field, and thus it is rarely conducted in atomistic ML. 
We believe there is great potential to explore uncertainty recalibration techniques tailored for atomistic ML models, improving their calibration and predictive reliability.

A third pressing challenge is related to the scarcity of UP techniques. 
Although UQ is reasonably investigated, UP receives far less attention despite its importance in chemical and materials modeling.
We suspect that this is partly due to the complexity of integrating UQ methods with physics-based simulations, which often involve solving differential equations and dealing with complex boundary conditions.
A promising direction to tackle the challenge is to develop fully automatic differentiable simulation approaches. 
These approaches combine automatic differentiation with physics-based simulations, enabling end-to-end differentiation of entire simulation pipelines. 
As a result, they will allow for the seamless propagation of uncertainties from the ML models to the quantities of interest.

If the existing challenges in UQ and UP can be overcome, we foresee substantial opportunities to accelerate the adoption and development of reliable and robust atomistic ML models. 
This will enable the exploration of complex chemical and materials systems with quantified uncertainties, ultimately leading to more informed decision-making and accelerated discovery.

\appendix
\section{Abbreviations}
\label{sec:abbr}
 
\begin{table}[H]
\setlength{\tabcolsep}{8pt}
\begin{tabular}{ll}
    CDF   & Cumulative distribution function \\
    DFT   & Density functional theory \\
    ENCE  & Expected normalized calibration error \\ 
    i.i.d.\ & Independent and identically distributed \\ 
    MAE  & Mean absolute error \\
    MC   & Monte Carlo \\ 
    MCMC & Markov chain Monte Carlo \\ 
    MD   & Molecular dynamics \\
    ML   & Machine learning \\
    MLE  & Maximum likelihood estimation \\
    NLL  & Negative log-likelihood \\
    NN   & Neural network \\ 
    QoI  & Quantity of interest \\
    RMSE & Root-mean-square error \\
    UQ   & Uncertainty quantification \\
    UP   & Uncertainty propagation \\
\end{tabular}
\end{table}

\section*{Conflicts of interest}
There are no conflicts of interest to declare.

\section*{Acknowledgements}
This work is supported by the National Science Foundation under Grant No.\ 2316667 and the startup funds from the Presidential Frontier Faculty Program at the University of Houston.

\bibliography{main.bib}

\end{document}